\documentstyle[11pt,aaspp4]{article}

\def\pdot{\hbox{$\buildrel . \over {\bf p}$}\ }
\def\aap{Astron. Astrophys.\ }

\def\PLB{{Phys. Lett.}  B}
\def\PRL{{Phys. Rev. Lett.}\ }
\def\PRD{{Phys. Rev.} D}
\def\PRC{{Phys. Rev.} C}

\def\vep{\varepsilon}

\def\beq{\begin{equation}}
\def\eeq{\end{equation}}
\def\beqa{\begin{eqnarray}}
\def\eeqa{\end{eqnarray}}

\def\etal{{\it et~al.\,}}

\newcommand{\lte}{\lower 0.5ex\hbox{${}\buildrel<\over\sim{}$}}
\newcommand{\eqn}[1]{(\ref{#1})}
\newcommand{\p}{\partial}
\newcommand{\inu}{I_\nu}
\newcommand{\hnu}{H_\nu}
\newcommand{\jnu}{J_\nu}
\newcommand{\pnu}{P_\nu}

\newcommand{\Unu}{U_\nu}
\newcommand{\vnu}{V_\nu}

\newcommand{\aenu}{\bar{\nu}_e}
\newcommand{\enu}{\nu_e}
\newcommand{\sig}{\sigma}
\newcommand{\f}{{\cal{F}}}

\newcommand{\pr}{^\prime}

\newcommand{\righta}{\rightarrow}

\lefthead{Burrows \etal}
\righthead{Neutrino Transport}

\begin{document}

\title{A New Algorithm for Supernova Neutrino Transport \\ and Some Applications}
\author{Adam Burrows\altaffilmark{1}, Timothy Young\altaffilmark{1}, 
Philip Pinto\altaffilmark{1,2}, Ron Eastman\altaffilmark{2,3}, and Todd A. Thompson\altaffilmark{4}}
\begin {center}
aburrows,tyoung,ppinto@as.arizona.edu,
reastman@llnl.gov, thomp@physics.arizona.edu
\end {center}

\altaffiltext{1}{Department of Astronomy and Steward Observatory, 
                 The University of Arizona, Tucson, AZ \ 85721}
\altaffiltext{2}{Lawrence Livermore National Laboratory, Livermore, CA \ 94551} 
\altaffiltext{3}{Department of Astronomy and Astrophysics, University of California, 
                 Santa Cruz, CA 95064}
\altaffiltext{4}{Department of Physics, 
                 The University of Arizona, Tucson, AZ \ 85721}

\begin{abstract}

We have developed an implicit, multi--group, time--dependent, spherical neutrino transport code based on the Feautrier 
variables, the tangent--ray method, and accelerated ${\bf \Lambda}$ iteration.  The code achieves 
high angular resolution, is good to O($v/c$), is equivalent to a Boltzmann solver (without gravitational redshifts), 
and solves the transport equation at all optical depths with precision.  In this paper, we 
present our formulation of the relevant numerics and microphysics and explore 
protoneutron star atmospheres for snapshot post--bounce models. Our major 
focus is on spectra, neutrino--matter heating rates, Eddington factors, 
angular distributions, and phase--space occupancies.  In addition, we investigate
the influence on neutrino spectra and heating of final--state electron blocking, 
stimulated absorption, velocity terms in the transport equation, neutrino--nucleon
scattering asymmetry, and weak magnetism and recoil effects.   
Furthermore, we compare the emergent spectra and heating rates obtained using full transport
with those obtained using representative flux-limited transport formulations to
gauge their accuracy and viability.  Finally, we derive useful formulae
for the neutrino source strength due to nucleon--nucleon bremsstrahlung and determine 
bremsstrahlung's influence on the emergent $\nu_{\mu}$ and $\nu_{\tau}$ neutrino spectra. These studies
are in preparation for new calculations of spherically symmetric core--collapse supernovae, protoneutron star winds,
and neutrino signals.

\end{abstract}



\section{Introduction}

With core--collapse supernova explosions, Nature
has contrived an elegant means to create compact objects,
while at the same time seeding the galaxy with the elements of existence.
Neutrinos play a key role in the phenomena of collapse and explosion,
for not only are they produced in abundance at the high temperatures and densities
achieved in collapse, but they are weakly enough coupled to matter that they transport heat 
and leptons on a dynamically interesting timescale.  It is now thought that neutrino
heating of the protoneutron star mantle drives the supernova 
explosion (Colgate and White 1966; Bethe and Wilson 1985), 
but only after a post--bounce delay of 100's of milliseconds to one second. 
During this delay, the quasi--static accreting core, the protoneutron star bounded
by the stalled shock wave, radiates neutrinos of all species
and the net energy deposition in the semi--transparent ``gain region" behind 
the shock plays a pivotal role in ``igniting" the explosion.  However, 
the precise deposition rate depends upon the details of neutrino transfer
at low ``optical" depths, putting great demands upon the theoretical tools
employed to calculate the properties of the neutrino radiation fields.
The character of that radiation depends upon neutrino--matter opacities, neutrino production
source terms, and neutrino transport.  Over the years, neutrino transport theory
and the associated microphysics have reached a sophisticated level 
of refinement (Tubbs and Schramm 1975; Lichtenstadt {\it et al.} 1978; Bowers and Wilson 1982; 
Mayle, Wilson, and Schramm 1987; Bludman and Schinder 1988; 
Bruenn 1985; Janka 1991; Mezzacappa and Bruenn 1993a,b; Messer {\it et al.} 1998;
Yamada, Janka, and Suzuki 1999).  However, despite these efforts, recent progress in modeling supernovae,
and new insights gained into the character of multi--dimensional neutrino-driven explosions
(Herant {\it et al.} 1994; Burrows, Hayes, and Fryxell 1995; Janka and M\"uller 1996;
Mezzacappa {\it et al.} 1998), the supernova explosion problem is not solved in detail.
We know little about the dependence of the $^{56}$Ni yields on progentior mass and composition, 
the iron--peak nucleosynthesis,
the explosion energies, the nascent pulsar kicks, and the asymmetries and mixing in the explosion debris.         
Furthermore, we still do not know the duration of the post--bounce delay, nor the ensemble of 
possible neutrino signatures.

In the past, a variety of approximations to the full neutrino transport equations have been
employed in complex numerical codes meant to simulate stellar collapse and supernova explosions.
These compromises have been deemed necessary because of the severe CPU constraints
of such simulations, particularly when those simulations have been multi--dimensional 
(Herant {\it et al.} 1994; Burrows, Hayes, and Fryxell 1995; Janka and M\"uller 1996).
A variety of gray approaches, flux limiters, equilibrium assumptions, and approximations
to both neutrino source and redistribution terms have been employed,
sometimes to good effect.  However, given the marginality of the explosions thus far
obtained, the fact that there is as yet no unanimity among theorists concerning important
issues of principle (cf. Mezzacappa {\it et al.} 1998), and the manifest importance of neutrinos
in collapse phenomenology, a fresh look at neutrino transport and the relevant neutrino physics
is in order.  It is in that spirit that we have constructed an implicit, time--dependent, multi--group, multi--angle,
multi--species neutrino transfer code to simulate the neutrino radiation fields in stellar collapse and explosion.
This code embodies a different computational method from that used in the pioneering papers
by Bruenn (1985) and Mezzacappa (Mezzacappa and Bruenn 1993a,b), but in its use of Feautrier 
variables and the tangent--ray method it is quite in keeping with traditional photon 
transport and stellar atmospheres simulations (Mihalas 1980; Mihalas and Mihalas 1984).  
In this paper, we describe the basic algorithm,
discuss and derive the relevant neutrino microphysics, and present high--resolution (in energy, angle, and radius) 
results for representative post--bounce protoneutron star configurations.  Hence, for this first paper 
in our series on neutrino transport and microphysics we focus on precision neutrino ``atmospheres."  We present
the energy spectra, Eddington factors, angular distributions, phase space densities, and
neutrino--matter energy couplings.  We also derive or discuss the relevant neutrino physics, 
some of it new.  We calculate the background neutrino radiation fields for two snapshot models, 
one of which is from the work of Burrows, Hayes, and Fryxell (1995) 
representing the wind phase that follows explosion (Model W), 
and one of which (our Model BM) was kindly provided to us by Tony Mezzacappa and is from a 
multi--group, flux--limited diffusion simulation by Bruenn (Messer {\it et al.} 1998), 106 
milliseconds after the bounce of the core of a Weaver and Woosley (1995)
15 M$_{\odot}$ star.  These are meant to exemplify various protoneutron
star structures and phases for the purposes of a detailed scrutiny   
of the neutrino sector.  Consistent dynamical calculations will follow later 
in the series.

Neutrinos are the major signatures of the inner turmoil of the dense core of the
massive star and they carry away the binding energy of the young neutron star, 
a full 10\% of its mass--energy. The detection of collapse neutrinos,
their ``light curve'' and spectra, will allow us to follow
in real time the phenomena of stellar death and birth.
The supernova, SN1987A, provided a glimpse of what might be possible, but
it yielded only 19 events; we can expect the current generation of underground neutrino
telescopes to collect thousands of events from a galactic supernova.

In \S\ref{sec2}, we present the equations and physics of neutrino transport. 
In \S\ref{sec3} we describe our implementation of the Feautrier and tangent--ray schemes,
and follow this in \S\ref{sec4} with a discussion of accelerated ${\bf \Lambda}$ iteration and 
our approach to the implicit coupling of matter with neutrino radiation.  Section 5
contains a physical derivation of stimulated absorption and \S\ref{cross6}
summarizes the cross sections and source terms we employ for this study.  
We provide in \S\ref{bremsst} a derivation
of the single and pair neutrino rates and spectra due to nucleon--nucleon bremsstrahlung, a process
that can compete with pair annihilation as a source for $\nu_{\mu}$, $\bar{\nu}_{\mu}$, 
$\nu_{\tau}$, and $\bar{\nu}_{\tau}$ neutrinos and that to date has not been incorporated
into supernova codes.  (The consequences of bremsstrahlung for the
emergent $\nu_{\mu}$ spectra are presented in \S\ref{numu}.)   In \S\ref{ntr}, we 
present for generic protoneutron star configurations our basic results vis \`a vis
emergent energy spectra, luminosities, and energy deposition rates (including
that due to $\nu \bar{\nu}$ annihilation).  We also explore the dependence of the emergent spectra
and neutrino heating rates on blocking factors, weak magnetism and recoil, aberration and Doppler terms, and stimulated absorption. 
These terms/effects are frequently dropped in simpler schemes.
In \S\ref{limit}, we highlight the angular dependence of the radiation fields and the conceptual limitations
of flux limiters that ignore the angular dimension.  Moreover, we compare the 
emergent spectra and heating rates obtained using our
full transport code with those obtained using representative flux-limiter closures
in order to gauge the errors of such approximate schemes.

This paper contains a description of neutrino transfer, our numerical approach, and the new results
that flow from it.  It is also meant to summarize various useful formulae 
that others, as they approach the study of supernova neutrino 
radiation fields, might employ.  In assembling the rates and cross sections, we have borrowed 
from the investigations of Tubbs and Schramm (1975), Bruenn (1985), 
Janka (1991),  Mezzacappa and Bruenn (1993a,b,c), Schinder and Shapiro (1982), and Bowers and Wilson (1982),
but take full responsibility when we have chosen to deviate from the literature.

\section{Neutrino Transport Equations \label{sec2}}

We have constructed a radiation/hydrodynamic code that solves the 
three equations of hydrodynamics with the equations of multi--group radiative transfer
and composition.  The hydro code is a one--dimensional Lagrangean realization of
the explicit Piecewise--Parabolic Method (PPM) of Colella \& Woodward (1984) (Fryxell \etal 1991)
that is automatically conservative, second--order accurate in space and time, and 
employs a Riemann solver to handle shocks.  Radiation is coupled 
to the matter between the hydro updates in an
implicit, operator--split fashion, employing accelerated ${\bf \Lambda}$ iteration (ALI) to facilitate the
convergence both of the transport solution and of the temperature and 
composition changes due to transport. Since in this paper
we focus on the transport sector of the code and on precision neutrino atmospheres, we postpone
to a later paper a discussion of the full hydrodynamic technique 
and of time--dependent results in the stellar collapse and supernova context.  
Here we describe the radiation equations solved, the algorithm developed to solve them, 
and the philosophy behind our methods.  
In later sections, we explore the nature of the neutrino radiation fields in the 
post--bounce and protoneutron star contexts.  In addition, we study the
influence of various terms and physics on the emergent neutrino spectra and on the neutrino--matter coupling 
in the semi--transparent region between the neutrinospheres and the stalled shock.  Energy deposition
in this region is thought to be important in igniting and driving the supernova explosion.

Neutrino transport is not an esoteric subject apart from traditional radiative transfer.  
The same techniques developed for one particle type can be employed for another.
For all particles, the solution to the Boltzmann equation is sought.  What distinguishes
neutrino transport and transfer are the number of neutrino species (six), the particular microphysics
of the neutrino--matter interaction ({\it i.e.,} cross sections, sources), the Fermi statistics
of the neutrinos (manifest only in the collision term), and the fact 
that there is in principle a conserved lepton number.   Neutrino
oscillations can alter this, but given the particular neutrino masses and oscillation angles suggested
by the recent solar and atmospheric neutrino data (Suzuki 1998; Totsuka {\it et al.} 1998), 
oscillations might not dramatically affect supernova dynamics
or the neutrino fields in the core (Fuller \etal 1992).  (It should be borne in mind that
oscillations in the outer envelope of the progenitor massive star or between the supernova 
and the Earth may alter the signal detected in underground neutrino telescopes.)

There are a variety of ways of writing the transport equation 
for the specific intensity ($\inu$) of the radiation field (Mihalas 1980; Mihalas and Mihalas 1984).  
In principle, the Boltzmann equation and the transport equation are equivalent, though the former is 
written in terms of the invariant phase space density (${\cal F}_\nu$), related one--to--one to the
specific intensity through the identity:
\beq
\frac{I(\mu,\varepsilon)}{\varepsilon^3}=\frac{g}{h^3c^2}\f_\nu\,\,,
\label{invarianti}
\eeq
where $g=1$ for massless neutrinos, $g=2$ for photons, $\varepsilon$ is the particle's energy, and
the other symbols have their standard meanings.  Sometimes it is said that the Boltzmann equation 
is more general than the transport equation because it contains a \pdot term that for massless
particles corresponds to gravitational redshifts.  However, there is no reason to exclude such a term
from the transport equation and we will not engage in such distinctions. 

One form of the transport equation for $\inu$ in the comoving frame in 
spherically--symmetric geometry is:
$$\frac{1}{c}\frac{D\inu}{Dt}+\frac{\mu}{r^2}\frac{\p}{\p r}(r^2\inu)+
\frac{\p}{\p\mu}\left[(1-\mu^2)\left[\frac{1}{r}+\frac{\mu}{c}
\left(\frac{v}{r}-\frac{\p v}{\p r}\right)\right]\inu\right]-
\frac{\p}{\p \varepsilon}\left[\varepsilon\left[(1-\mu^2)
\frac{v}{cr}+\frac{\mu^2}{c}\frac{\p v}{\p r}\right]\inu\right]$$
\beq
+\left[(3-\mu^2)\frac{v}{cr}+\frac{1+\mu^2}{c}\frac{\p v}{\p
r}\right]\inu+{\cal A}_\nu=
\eta_\nu-\chi_\nu\inu+
\frac{\kappa_s}{4\pi}\int\Phi({\bf \Omega},{\bf \Omega\pr})\,I_\nu({\bf
\Omega\pr})\,d{\bf\Omega\pr}\,\, ,
\label{transport1}
\eeq
where
\beq
{\cal A}_\nu=\frac{a}{c^2}
\left[3\mu -(1-\mu^2)\frac{\p }{\p \mu}-\varepsilon\frac{\p }{\p
\varepsilon}\right]I_\nu\, ,
\label{accel}
\eeq
$a$ is the matter acceleration, $\mu=\cos\theta$,     
$\varepsilon$ is the neutrino energy, and
$\eta_\nu$ is the emissivity of the medium.  The subscript $\nu$ indicates
neutrino energy dependence and $\Phi$ is an angular phase function for neutrino
scattering into the beam.  This equation, good to O($v/c$) (where 
$v$ is the matter velocity and $c$ is the speed of light),
assumes azimuthal symmetry and contains the appropriate redshift, aberration, and advection
terms due to matter motion, angular redistribution due to 
scattering into the beam, scattering and absorption
out of the beam, and source terms.  Eq. (\ref{transport1}) does not include energy 
redistribution upon scattering, to be incorporated in a later version of the code.  The various terms represent 
the additions and subtractions from the beam, the entire 
equation representing conservation of energy and number.  
The microphysics and collision terms reside on the right--hand--side and 
the geometry, aberration, advection, and Doppler shift terms reside on the left.  

While eq. (\ref{transport1}) contains the relevant terms to O($v/c$), it is a bit awkward
to difference.  It is also a bit ugly and its various terms 
are not so cleanly distinguished by their physical roles.  Dropping the acceleration
term, we follow Eastman and Pinto (1993) and derive the form of
the transport equation we employ in this study.   
The equation, physically equivalent to the
Boltzmann equation (ignoring gravitational redshifts and the acceleration term) for an individual
neutrino species, is:

$$\frac{1}{c}\frac{D\inu}{Dt}+\mu\frac{\p \inu}{\p r} +
\frac{1-\mu^2}{r}(1-\beta Q \mu)\frac{\p \inu}{\p \mu}
+\frac{\beta}{r}\left(1+Q\mu^2\right)\left(3-\frac{\p}{\p
\rm{ln}\varepsilon}\right)I_\nu$$
\beq
=\eta_\nu-\chi_\nu\inu +
\frac{\kappa_s}{4\pi}\int\Phi({\bf \Omega},{\bf \Omega\pr})\,I_\nu({\bf
\Omega\pr})\,d{\bf\Omega\pr}\,\, ,
\label{east01}
\eeq
where $Q \equiv\p{\rm ln} v/\p{\rm ln} r-1$ and all other symbols
have their standard meanings.  $\Phi$ is a phase function for neutrino
scattering into the beam.
$\chi_\nu$ is the total extinction coefficient ($=\kappa_a+\kappa_s$), where $\kappa_a$ and $\kappa_s$
contain contributions from all absorption
and scattering processes, respectively:
\beq
\kappa_s=\sum_i \,n_i\,\sig^s_i
\hspace{.75cm}
{\rm and}
\hspace{.75cm}
\kappa_a=\sum_i\, n_i\,\sig^a_i\,\, .
\label{kappadef}
\eeq

Eq. (\ref{east01}) can be rewritten as the corresponding Boltzmann equation for $\f_\nu$: 
$$\frac{1}{c}\frac{D\f_\nu}{Dt}+\mu\frac{\p \f_\nu}{\p r} +
\frac{1-\mu^2}{r}(1-\beta Q \mu)\frac{\p \f_\nu}{\p \mu}
- \frac{\beta}{r}\left(1+Q\mu^2\right)\frac{\p \f_\nu}{\p
\rm{ln}\varepsilon}$$
\beq
=\frac{h^3c^2}{g}\left(\frac{\eta_\nu}{\varepsilon^3}\right)-\chi_\nu\f_\nu+\frac{\kappa_s}{4\pi}\int\Phi({\bf
\Omega},{\bf \Omega\pr})\,\f_\nu({\bf \Omega\pr})\,d\Omega\pr \, ,
\label{invartrans1}
\eeq
which can be mapped directly, term--by--term, into the Boltzmann equation 
employed by Messer \etal (1998) in their recent work on Boltzmann neutrino transfer. 
Eq. (\ref{invartrans1}) is the most useful form of the transport equation when studying
it using the method of characteristics.

For neutrinos, the phase function for a scattering process $i$ is well approximated
(except for $\nu$--$e^-$ scattering) by
\beq
\Phi_i({\bf \Omega},{\bf \Omega\pr})=\Phi_i({\bf \Omega \cdot \Omega\pr})=
\left(1+\delta_i\,{\bf \Omega \cdot
\Omega\pr}\right)=\left(1+\delta_i\mu\right)\,\, ,
\label{phase}
\eeq
where $\delta_i$ is a constant specific to each scattering process and here $\mu$
is the angle between the incident and outgoing neutrinos.
Hence, we can write the differential cross section for a scattering
process $i$ in terms of the total scattering cross section:
\beq
\frac{d\sig^s_i}{d\Omega}=\frac{\sig^s_i}{4\pi}(1+\delta_i\mu)\,\, .
\label{diffcs}
\eeq

Subsequently, we drop the superscript $s$.
Eq. (\ref{phase}) implies that the angular redistribution term in eq.
(\ref{east01}) becomes
\beq
\kappa_s J_\nu+\frac{\kappa_s\delta_T}{4\pi}{\bf \Omega\cdot F}_\nu\,\, ,
\label{phase1}
\eeq
where
\beq
\delta_T=\frac{\sum_i \,n_i\,\sig_i\, \delta_i}{\sum_i\, n_i\, \sig_i}\,\,
.
\eeq
$F_\nu$ in eq. (\ref{phase1}) is the neutrino flux and $J_\nu$ is the
zeroth moment defined by
\beq
J_\nu=\frac{1}{2}\int_{-1}^1\,I_\nu\,d\mu=\frac{c}{4\pi}E_\nu\,\, ,
\label{0moment}
\eeq
where $E_\nu$ is the neutrino energy density.

Integrating eq. (\ref{east01}) and ${\bf \Omega}\,\times$ eq.
(\ref{east01}) over $d{\bf \Omega}$ yields the zeroth and first moment
equations, respectively:
\beq
\frac{1}{c}\frac{D\jnu}{Dt}+\frac{1}{r^2}\frac{\p}{\p r}(r^2 H_\nu) -
\frac{\beta Q}{r}(3P_\nu-\jnu) +
\frac{\beta}{r}\left(3-\frac{\p}{\p \rm{ln}
\varepsilon}\right)(\jnu+QP_\nu)
=\kappa_a^*(B_\nu-J_\nu)
\label{east02}
\eeq
and
$$\frac{1}{c}\frac{D\hnu}{Dt}+
\frac{\p P_\nu}{\p r} +\frac{3P_\nu-\jnu}{r} -
\frac{\beta Q}{r}(4N_\nu-2H_\nu)+
\frac{\beta}{r}\left(3-\frac{\p}{\p \rm{ln}
\varepsilon}\right)(\hnu+QN_\nu)$$
\beq
=-\left(\kappa_a^*+\kappa_s-\frac{1}{3}\kappa_s\delta_T\right)H_\nu=
-\left(\kappa_a^*+\kappa_{tr}\right)H_\nu\,\, ,
\label{east03}
\eeq
where $H_\nu$, $P_\nu$, and $N_\nu$ are the first, second, and third angular
moments given by
\beq
H_\nu=\frac{1}{2}\int_{-1}^1 \mu\,I_\nu\,d\mu=\frac{1}{4\pi}F_\nu\,\, ,
\eeq
\beq
P_\nu=\frac{1}{2}\int_{-1}^1 \mu^2\,I_\nu\,d\mu\,\, ,
\eeq
and
\beq
N_\nu=\frac{1}{2}\int_{-1}^1 \mu^3\,I_\nu\,d\mu\,\, .
\eeq
$B_\nu$ is the equilibrium (black body) spectral energy density times $\frac{c}{4\pi}$  
and $\kappa_a^*$ includes the correction for stimulated absorption (see \S \ref{stimabs}).  
$\kappa_{tr}$ in eq. (\ref{east03}) is the total transport extinction coefficient 
and is defined in terms of the individual transport
cross sections as $\kappa_{tr}=\sum_i\, n_i\,\sig_i^{tr}$.  For a
particular scattering process $i$,
\beq
\sig^{tr}_i=\int\frac{d\sig_i}{d\Omega}(1-\mu)\,d\Omega
=\sig_i\left(1-\frac{1}{3}\delta_i\right)\,\,.
\label{transcs}
\eeq

Note that $\delta_p$ and $\delta_n$, the $\delta_i$s for neutrino-nucleon scattering, 
are negative ($\delta_p\sim-0.2$ and $\delta_n\sim-0.1$)
and, hence, that these processes are backward--peaked.
The fact that $\delta_p$ and $\delta_n$ are negative and, as a consequence, that
$\sig^{tr}_i$ is greater than $\sigma_i$ increases the neutrino--matter energy coupling rate
for a given neutrino flux in the semi--transparent region.  This increase in inverse
flux factor ($J_\nu/H_\nu$) is but one effect that can be studied with the
transport tools we have developed and are developing.

Integrating eq. (\ref{east02}) over energy, we obtain the neutrino energy
equation:
\beq
\frac{DE}{Dt}+\frac{1}{r^2}\frac{\p}{\p r}(r^2F)
-\frac{v}{r}(3p-1)QE+4\frac{v}{r}(1+Qp)E
=4\pi\int_0^\infty\kappa_a^*(B_\nu-J_\nu)\,d\varepsilon\,\, ,
\label{energy}
\eeq
where $E$ and $F$ are the integrated neutrino energy density and flux,
respectively. $p$ is the energy--integrated Eddington factor, where
$p_\nu={P_\nu}/{J_\nu}$.  
The sums for all neutrino species of the negative of the right--hand--side of 
eq. (\ref{energy}) and the negative of the $\varepsilon_\nu$ integral 
of the right--hand--side of eq. (\ref{east03}) are the 
energy and momentum source terms in the matter equations. 
The two equations for the rate of change of the electron fraction ($Y_e$) 
due to $e^-$/$e^+$/$\nu_e$/$\bar{\nu}_e$ capture are:

\beq
\rho{\cal N_A}\frac{DY_e}{Dt}_\pm=\pm4\pi\int_0^\infty\kappa_a^*(B_\nu-J_\nu)_{\pm}\,\frac{d\varepsilon}{\varepsilon}\, ,
\label{yeeq}
\eeq
where the $-$ sign is for the $\nu_e$ equation and the $+$ sign is for the $\bar{\nu}_e$ equation.  
Eqs. (\ref{east01}), (\ref{east02}), and (\ref{east03}) for each neutrino species, along with eqs. (\ref{yeeq}),
are the basic neutrino transport equations that we solve. 
$\rho$ is the mass density
and $\cal N_A$ is Avogadro's number. 

\section{Method of Solution: Feautrier and Tangent--Ray Algorithm\label{sec3}}

We solve the moment eqs. (\ref{east02}) and (\ref{east03}) implicitly for $J_\nu$ and
$H_\nu$ by backwards differencing in time the quantities $J_\nu/\rho^{4/3}$ and
$H_\nu/\rho^{2/3}$, backwards differencing in $\ln\varepsilon$ 
(according to the slope of the characteristic), and combining
the spatially differenced equations into
one equation for $J_\nu$ which is 2nd--order accurate in $r$.
Standard matrix inversion techniques are employed to obtain $J_\nu$,
from which $H_\nu$ is derived using eq. (\ref{east03}).
This equation is manifestly Lagrangean and by solving it 
the advective derivative is included automatically.
$J_\nu/\rho^{4/3}$ and $H_\nu/\rho^{2/3}$ are the natural combinations
for adiabatic compression or expansion of a relativistic gas.

Since eqs. (\ref{east02}) and (\ref{east03}) contain the higher--order angular
moments $P_\nu$ and $N_\nu$, closure relations are needed.  These are
obtained from a formal integration of the full transport equation (\ref{east01}), written in terms
of the Feautrier variables:
\beq
\Unu(\mu)=\frac{1}{2}(\inu(\mu)+\inu(-\mu))
\hspace{1cm}{\rm{and}}\hspace{1cm}
\vnu(\mu)=\frac{1}{2}(\inu(\mu)-\inu(-\mu))\,\, ,
\eeq
where $\mu$ ranges from 0 to 1.

In the isotropic scattering limit ($\delta_T=0$), these equations for $\Unu$ and $\vnu$ are:
\beq
\frac{1}{c}\frac{D\Unu}{Dt}+\mu\frac{\p \vnu}{\p r}-
\frac{1-\mu^2}{r}\beta Q\mu\frac{\p U_\nu}{\p \mu}+
\frac{\beta}{r}
\left(3-\frac{\p}{\p \rm{ln} \varepsilon}\right)(1+Q\mu^2)\Unu
=\eta_\nu-\chi_\nu U_\nu+\kappa_s\jnu
\label{UNU}
\eeq
and
\beq
\frac{1}{c}\frac{D\vnu}{Dt}+\mu\frac{\p \Unu}{\p r}-
\frac{1-\mu^2}{r}\beta Q\mu\frac{\p \vnu}{\p \mu}+
\frac{\beta}{r}
\left(3-\frac{\p}{\p \rm{ln} \varepsilon}\right)(1+Q\mu^2)\vnu=
-\chi_\nu\vnu\,\, .
\label{VNU}
\eeq
From the solution of eqs. (\ref{UNU}) and (\ref{VNU}), we obtain the 
full radiation field and the higher--order moments that are then used in eqs. (\ref{east02}) 
and (\ref{east03}) for $J_\nu$ and $H_\nu$.  Since eqs. (\ref{UNU}) and (\ref{VNU})
require the lower--order moment $J_\nu$ (and in principle $H_\nu$, {\it cf.} eq. \ref{phase1}),  
we iterate this system until we obtain a converged and consistent global solution.  
Simultaneously, we calculate the ${\bf \Lambda}$ operator that maps $S_\nu$,
the source function, 
onto $J_\nu$ and employ accelerated ${\bf \Lambda}$ iteration 
(Cannon 1973a,b; Scharmer 1981; Olson, Auer, and Buchler 1986; 
Eastman and Pinto 1993) to speed the convergence of the temperature and composition updates.
Independent of the total optical depth, this generally requires no more 
than 2 to 3 iterations to obtain an accuracy of a part in $10^{6}$.
To maintain stability and reflect the density character of $\Unu(\mu)$ and
the flux character of $\vnu(\mu)$, we stagger the $\Unu(\mu)$ and $\vnu(\mu)$ meshes 
with respect to one another by half a zone.

It may seem that by solving the moment equations separately and iterating with the solution of the 
transport equation itself and by not focusing simply
on the solution of eqs. (\ref{UNU}) and (\ref{VNU}) or eq. (\ref{east01}) that
we are doing more than is necessary.  An advantage of solving the moment equations is that they can be
differenced to automatically conserve energy.  However, enforcing energy conservation by construction does not guarantee
that the solution obtained is the correct one.  In fact, it is standard 
with many ``non-conservative'' hydrodynamic codes that do not 
difference the equations to conserve energy by construction to employ the degree to which
energy is in fact conserved with time to assess their accuracy.  
We have chosen to retain the automatic energy-conserving feature,
but in dynamical calculations we use electron lepton number conservation the global code check.  By not
differencing the equations to automatically conserve lepton number, since the equations we difference
certainly do conserve lepton number, this is a useful approach.  
This is akin to employing entropy conservation in adiabatic flow as a check of a hydro code 
that is forced to conserve energy automatically by the particular differencing scheme 
employed.  However, for the stationary atmospheres calculations we present here, 
for which the time derviative is zero, this is a moot point.  

The advantage of calculating $U_\nu$ and $V_\nu$ instead of $I_\nu$ is that
equations~\eqn{UNU} and \eqn{VNU} can be differenced in such a way that
$V_\nu$ will go accurately to $3\mu \partial S_\nu/\partial \tau_\nu$
in the large--optical--depth, diffusion limit, and still remain accurate
in the optically--thin, free--streaming limit.  Schemes based directly on
eq.~\eqn{transport1} or \eqn{east01} have the
correct large--optical--depth behavior for $U_\nu$, {\it i.e.} $U_\nu=S_\nu$,
but have round--off trouble computing $V_\nu$, which is important if the only
estimate of the flux comes from integrating $\mu V_\nu$ over angle
(Larsen, Morel, and Miller 1987).
Typically in such codes above a certain optical
depth the diffusion limit is assumed and $V_\nu$ is set equal to $3\mu
\partial S_\nu/\partial \tau_\nu$.

Numerical methods which solve for the spatial variation of a
specific-intensity-like-variable ({\it e.g.}, I$[\mu,r,E,t]$), such as all discrete
ordinate transport methods, suffer from the problem that at large
optical depth, the flux is not well determined 
(Morel, Wareing, and Smith 1996). This is a well-known
problem, and the principle motivation for switching to Feautrier
variables and to full range moment methods, in which $J$ and $H$ are solved
for directly. Any S$_N$ method has a spatial truncation error which is
proportional to $(\Delta\tau)^k$, where $k$ depends on the spatial discretization
scheme. As $\Delta\tau$ grows, the error in the flux, which is proportional to
$(\Delta\tau)^k$, also grows.  For Feautrier variables, on the other hand,
finite difference systems of equations can be derived which are at least 
2'nd-order accurate in $\tau$.  These give an accurate representation of the
radiation field in the free streaming limit and go naturally
over to the diffusion limit when $\Delta\tau$ is large.

To solve eqs. (\ref{UNU}) and (\ref{VNU}), we employ the tangent--ray method (Schinder and Bludman 1989;
Mihalas and Mihalas 1984).
At a reference radial zone, tangents are constructed to each of the interior zones.  The angles
of the tangent rays to the normal at the reference zone define the angular
grid at that zone on which the angular integrations are performed.  Eqs. (\ref{UNU}) and (\ref{VNU})
are integrated along each tangent ray.  If there are
$nx$ radial zones, the radiation field at the outer zone is resolved with $nx-1$ angles;
as you move inward the number of angles employed decreases linearly.  Hence, if
there are 100 radial zones, there are as many as 99 angular bins.  With reasonable radial
gridding, this approach can provide exquisite angular resolution, particularly
for forward--peaked radiation fields, but at the cost of
increased computational overhead.  For instance, we have tested our implementation
of the tangent--ray method with the Kosirev (1934) (spherical Milne) problem for
which the absorptive opacity is assumed to be a power law in radius ($\kappa = 1/r^n$).  For a variety
of integer power laws ({\it e.g.}, $n=1.1,1.3,1.5,2,3,4$), with from 100 to 500 radial zones,
the tangent-ray method is superior to many implementations of the 
discrete ordinate method (Schinder and Bludman 1989). 
However, care must be taken to avoid purely
geometrical zoning, for which $r_{n+1} = \alpha r_{n}$, since such zoning biases
the angular binning in a systematic way.  The result can be that the Eddington and flux factors
asymptote at infinity to between 0.96 and 0.98, and not to 1.0.  However, purely geometrical zoning is
easily avoided in real calculations.  
Note that the increase in angular resolution with radius which comes naturally 
from this procedure is quite appropriate to spherical symmetry.  Note also
that the tangent-ray method is in some sense automatically 
adaptive for moving grids.   At $r=0$, the radiation
field is by definition symmetric and needs no angular resolution.  As $r$ becomes large, a small 
bright central source is increasingly finely resolved.

For dynamical, time--dependent calculations,
solving eqs. (\ref{UNU}) and (\ref{VNU}) by the tangent-ray method at each timestep 
can be time-consuming, but manageable.  Fortunately, as long as the 2'nd and 3'rd angular moments do not
change quickly, one need not solve eqs. (\ref{UNU}) and (\ref{VNU}) at every timestep.
Frequently, the solution to the moment equations (\ref{east02}) and (\ref{east03}) with
previous values of $p_\nu$ and $g_\nu$ ($=N_\nu/H_\nu$) can be quite accurate.
In fact, this is most often the case, since the neutrino radiation fields rarely change on timescales
shorter than $\sim$0.1 milliseconds, whereas, due to the explicit nature of the hydro portion
of the code, the Courant timesteps are often near one microsecond. 
Hence, during much of the pre--explosion delay and protoneutron 
star phases, as well as during much of the core collapse phase,
it is quite legitimate to solve for the Eddington factors only every few steps.
An exception is during very dynamical phases such as shock breakout through the neutrinospheres.

Currently, there are two approximations in our algorithm for solving the transport equations
\eqn{UNU} and \eqn{VNU}. The first is that in calculating the radiation angular moments 
we assume that scattering into the beam is isotropic, while maintaining the
correct transport cross section in the $H_\nu$ moment eq. (\ref{east03}).
To approximately incorporate the effects of anisotropic scattering, 
we employ the transport cross sections, as described and discussed above.    
Straightforward methods for solving the fully anisotropic problem using the Feautrier 
variables and the approaches outlined in this paper will be described elsewhere.
The second approximation involves the assumption of linearity of the characteristics.
The 2'nd term in both eqs. \eqn{UNU} and \eqn{VNU} which
is proportional to ${\partial/ \partial \mu}$ comes from the aberration
experienced in going from the local to a nearby rest frame. 
The characteristics are not perfectly straight, which can make
the calculation more difficult.
One cannot simply integrate along a
straight--line impact ray. However, these terms are often
insignificant because we require only an estimate
of $U_\nu$ and $V_\nu$ and are using them to
compute only the closure coefficients, $p_\nu$ and $g_\nu$. Rather than just
ignore these two terms, we have substituted
\begin{equation}
{1-\mu^2\over r}\beta Q\mu{\partial U_\nu\over \partial \mu} \rightarrow
{3\mu^2-1\over r}\beta Q U_\nu
\label{Usubs}
\end{equation}
in equation \eqn{UNU}
and
\begin{equation}
{1-\mu^2\over r}\beta Q\mu{\partial V_\nu\over \partial \mu} \rightarrow
{4\mu^2-2\over r}\beta Q\mu V_\nu
\label{Vsubs}
\end{equation}
in equation \eqn{VNU}.  The substitution in eq. (\ref{Usubs}) is derived by
integrating the left--hand--side by parts and the substitution in eq. (\ref{Vsubs}) 
is derived by integrating $\mu$ times the left--hand--side by parts.
Importantly, the two terms in \eqn{Usubs} integrate to the
same thing and $\mu$ times the two terms in eq. \eqn{Vsubs} integrate
to the same thing. Therefore, both modified equations reduce to the
energy and momentum conservation equations.

In sum, we solve two coupled moment equations for the mean intensity and flux of
the radiation field. These are the fundamental results of the transport
calculation. They are solved by an Eddington factor iteration wherein a
set of angle-dependent equations consistent with the moment equations are
solved for the intensity given a constant source function, and this
intensity is used to determine the closure factors in the moment
solution.

\section{Implicit Coupling to Matter and Accelerated $\Lambda$ Iteration\label{sec4}}

Though we are not highlighting in this paper time--dependent calculations,
we think it useful to include a discussion of the technique we employ to couple matter with neutrinos. 
This is done implicitly in operator--split fashion, after each hydro update.   For each neutrino
species, the scattering and absorption opacities and the emissivities are
calculated and fed into the transport solver.  A fully converged solution
of the transport equations is obtained and this is used to calculate the various
terms needed for the implicit update of the temperature and $Y_e$ due to transport. 
In particular, the derivatives with respect to temperature and $Y_e$ of the right--hand--sides of eqs. 
(\ref{energy}) and (\ref{yeeq}) are calculated.  For the implicit temperature update at 
each radial zone, $i$, a backward--differenced matter energy equation like:
\beq
\rho C_V\frac{T^{k+1}_i-T^{k}_i+\Delta T_i}{\Delta t}=-4\pi\int_0^\infty(\eta^{\prime}-\kappa_a^*J_\nu)\,d\varepsilon
-4\pi{\Delta T_i}\int_0^\infty\Bigl(\frac{\p\eta^{\prime}}{\p T} - 
\frac{\p\kappa_a^*}{\p T}J_\nu - \kappa_a^*\frac{\p J_\nu}{\p T}\Bigr)\,d\varepsilon
\label{couple}
\eeq
is constructed, where ${\Delta T_i}$ is the temperature change between iterations, $T^{k+1}_i$ is
the new temperature, $T^{k}_i$ is the old temperature, $C_V$ is the specific heat, $\eta^{\prime}$ is not 
corrected for final--state neutrino blocking (\S\ref{stimabs}), and $\Delta t$ is the timestep.
(In fact, the matter energy is a function of both $T$ and $Y_e$ and there is an extra term
in the temperature update equation to account for the entropy change due to the $Y_e$ composition
change. That term has been dropped here for clarity, but not in the computations.)

The subtlety with eq. (\ref{couple}) lies in the $\frac{\p J_\nu}{\p T}$ term.  In general,
$J_i$ equals ${\bf \Lambda}_{ij}S_j$ at each frequency or energy, where the 
${\bf \Lambda}$ operator is a matrix coupling different zones. 
Hence, eq. (\ref{couple}) is a matrix equation with 

\beq
\frac{\p J_i}{\p T} = {\bf \Lambda}_{ij}\frac{\p S_j}{\p T}\, .
\label{jlambdas}
\eeq
For simplicity in eq. (\ref{couple}), we have dropped in eq. (\ref{jlambdas}) the ${\bf \Gamma}$ operator that couples energy groups.
Though we have the option in the code of calculating the full ${\bf \Lambda}$ matrix, we use only the diagonal and the two
adjacent off--diagonals.  It is this truncated tridiagonal ${\bf \Lambda}$ operator that we actually employ.  

Since the $S_\nu$ we use in eq. (\ref{jlambdas}), perhaps a bit idiosyncratically, 
equals $\eta^{\prime}/\kappa^*$, $\frac{\p S}{\p T}$ is given by 
\beq
\frac{\p S_i}{\p T} = \Bigl(\frac{\p \eta^{\prime}_{i}}{\p T} -
S_i\frac{\p\kappa^*_{i}}{\p T}\Bigr)/ \kappa^*_{i}\, ,
\label{diagons}
\eeq
and this is employed to derive:
\beq
\frac{\p J_i}{\p T}={\bf \Lambda}_{i,i+1}\frac{\p S_{i+1}}{\p T} + {\bf \Lambda}_{i,i-1}\frac{\p S_{i-1}}{\p T} + 
{\bf \Lambda}_{i,i}\frac{\p S_i}{\p T}\, .  
\label{djdtdia}
\eeq
Plugging eq. (\ref{djdtdia}) into eq. (\ref{couple}), we solve for $\Delta T_i$ 
by inverting the tridiagonal matrix.  For the $\nu_e$ and $\bar{\nu}_e$ species, a similar
procedure is followed to obtain $\Delta Y_e$ from eq. (\ref{yeeq}).  Note that the integral over
neutrino energy is performed before the $T$ and $Y_e$ updates, 
which are not attempted for each energy group individually.

Once $\Delta T_i$ and $\Delta Y^{i}_{e}$ are obtained, $T^{k+1}_i$ is set equal 
to $T^{k}_i$ + $\Delta T_i$ and $Y^{k+1,i}_e$ is set equal to $Y^{k,i}_e + \Delta Y^{i}_e$.  We then
loop back to obtain a new transport solution with the new temperature and $Y_e$. 
This procedure is iterated until $\Delta T_i/T_i$ and $\Delta Y^{i}_e/Y^{i}_e$ are
suitably small (normally $10^{-6}$) for all zones, at which time we are left with a completely 
consistent set of $J_\nu$, $H_\nu$, $U_\nu$, $V_\nu$, $T$, and $Y_e$\,.   
$T^{k}_i$ and $Y^{i}_e$ are not changed during the iteration.  The total number of iterations
varies between 1 and 7, the latter only when $Y_e$ is changing fast in either the $\nu_e$ or the $\bar{\nu}_e$ modules.
The various neutrino fluids are updated in series, not in parallel and we generally follow three
species: $\nu_e$, $\bar{\nu}_e$, and ``$\nu_\mu$,'' the latter representing the 
sum of $\nu_\mu$, $\bar{\nu}_\mu$, $\nu_\tau$, and $\bar{\nu}_\tau$ neutrinos. 
Bunching these four neutrino species into one assumes that their cross sections 
and source terms are identical, which technically is false, but quantitatively reasonable).
Note that to achieve stable iteration, it is essential for the derivatives in eq. (\ref{couple}) to be accurate.
Among other things, this requires good derviatives of $\hat{\mu}$ ($= \mu_{n} - \mu_{p}$) with respect to $T$ and $Y_e$.
Analytic derivatives are preferred, but numerical derivatives for most quantities seem to work.

As stated previously, to achieve rapid convergence of the transport iteration we employ ALI, accelerated
${\bf \Lambda}$ iteration (Cannon 1973a,b; Scharmer 1981; Olson, Auer, and Buchler 1986).  
This entails an approximation that improves during the iteration.
In particular, we use 

\beq
{J}^{k+1}_\nu = {J}^k_\nu + {\bf \Lambda}^{k}\bigl({S}^{k+1}_\nu - {S}^{k}_\nu\bigr)\, ,
\label{approxlam}
\eeq
where ${\bf \Lambda}^{k}$ is the retarded ${\bf \Lambda}$ operator. 
We use only its diagonal and first off--diagonal terms.  This procedure  
accelerates and stabilizes the iteration, even if the optical depth
is large and the scattering albedo is high. Note that one cannot iterate
on the full ${\bf \Lambda}$ matrix (the inverse of the matrix
representation of the finite-difference transport equations) because its
eigenvalues are very close to the unit circle and the iteration stabilizes
instead of converging. Subtracting off a piece of the ${\bf \Lambda}$ matrix and
lagging the iteration of that piece allows the iteration to converge much
more rapidly.

\section{Stimulated Absorption \label{stimabs}}

The concept of stimulated emission for photons is well understood and studied, but
the corresponding concept of stimulated {\it absorption} for neutrinos is not so
well appreciated.  This may be because its simple origin in Fermi blocking and the
Pauli exclusion principle in the context of {\it net} emission is not often explained.  
The {\it net} emission of a neutrino is simply the difference between the emissivity and
the absorption of the medium:
\beq
{\cal {J}}_{net}=\eta_\nu-\kappa_aI_\nu\,\, .
\label{netemission}
\eeq
All absorption processes involving fermions will be inhibited by Pauli blocking due to
final--state occupancy.  Hence, $\eta_\nu$ in eqs. (\ref{netemission}) and (\ref{east01}) includes a blocking term,
$(1-\f_{\nu})$ (Bruenn 1985). $\f_{\nu}$ is the invariant distribution function
for the neutrino, whether or not it is in chemical equilibrium. 

We can derive stimulated absorption using Fermi's Golden rule.  For example, the net collision term
for the process, $\nu_en\leftrightarrow e^-p$, is:
\beqa
{\cal C}_{\nu_e n\leftrightarrow e^- p}&=&
\int\frac{d^3\vec{p}_{\nu_e}}{(2\pi)^3 }
\int\frac{d^3\vec{p}_n}{(2\pi)^3 }
\int\frac{d^3\vec{p}_p}{(2\pi)^3 }
\int\frac{d^3\vec{p}_e}{(2\pi)^3 }\,
\left(\sum_{s}{|{\cal{M}}|}^2\right) \nonumber \\
&& \nonumber \\
&\times&
\Xi(\nu_en\leftrightarrow e^-p)\,\,(2\pi)^4\,\delta^4({\bf p}_{\nu_e}+{\bf
p}_n-{\bf p}_p-{\bf p}_e)
\,\, ,
\label{rate01}
\eeqa
where ${\bf p}$ is a four-vector and
\beq
\Xi(\nu_en\leftrightarrow
e^-p)=\f_{\nu_e}\f_n(1-\f_e)(1-\f_p)-\f_e\f_p(1-\f_n)(1-\f_{\nu_e})\,\, .
\label{blocks}
\eeq
The final--state blocking terms in eq. (\ref{blocks}) are manifest, in particular that for the $\nu_e$ neutrino.
Algebraic manipulations convert $\Xi(\nu_en\leftrightarrow e^-p)$ in eq. (\ref{blocks}) into:
\beqa
\Xi(\nu_en\leftrightarrow e^-p)&=&
\f_n(1-\f_e)(1-\f_p)\left[\frac{\f_{\nu_e}\pr}{1-\f_{\nu_e}\pr}(1-\f_{\nu_e})-\f_{\nu_e}\right]
\nonumber \\
&=&\frac{\f_n(1-\f_e)(1-\f_p)}{1-\f_{\nu_e}\pr}\left[\f_{\nu_e}\pr-\f_{\nu_e}\right]\,\, ,
\label{kirch03}
\eeqa
where
\beq
\f_{\nu_e}\pr=[e^{(\varepsilon_{\nu_e}-(\mu_e-\hat{\mu}))\beta}+1]^{-1} 
\label{equileq}
\eeq
is an equilibrium distribution function for the $\nu_e$ neutrino 
and it has been assumed that only the electron, proton, and neutron
are in thermal equilibrium.  Note that in $\f_{\nu_e}\pr$
there is no explicit reference to a neutrino chemical potential, though 
of course in beta equilibrium it is equal to $\mu_e-\hat{\mu}$. 
There is no need to construct or 
refer to a neutrino chemical potential in neutrino transfer.

Using eq. (\ref{invarianti}), we see that eq. (\ref{kirch03}) naturally leads to:
\beq
{\cal {J}}_{net}=\frac{\kappa_a}{1-\f\pr_\nu}\left(B_\nu-I_\nu\right)=\kappa_a^*(B_\nu-I_\nu)\,\, .
\label{kirch}
\eeq
This is akin to the right-hand-side of eq. (\ref{east02}).   
If neutrinos were bosons, we would have found a (${1+\f\pr_\nu}$) in the denominator, but the
form of eq. (\ref{kirch}) in which $I_\nu$ is manifestly driven to $B_\nu$, the equilibrium
intensity, would have been retained.  From eqs. (\ref{kirch03}) and (\ref{kirch}), we
see that the stimulated absorption correction to $\kappa_a$ is $1/(1-\f\pr_\nu)$. 
If we want to retain the form of the collision term as expressed in eqs. (\ref{netemission})
or (\ref{east01}), the physics is unaltered and stimulated absorption
is not needed as a concept, as long as $\eta_\nu$ in eq. (\ref{east01}) 
contains the neutrino Pauli blocking term, $(1-\f_{\nu})$, without the prime.
However, by writing the collision term in the form of eq. (\ref{kirch}), with $\kappa_a$ corrected for stimulated
absorption, we have a net source term that clearly drives $I_\nu$ to equilibrium.  The timescale
is $1/c\kappa_a^*\,$.   Though the derivation of the stimulated absorption correction
we have provided here is for the $\nu_en\leftrightarrow e^-p$ process, this correction is quite general and
applies to all neutrino absorption opacities.

Kirchhoff's Law, expressing detailed balance, is:
\beq
\kappa_a = \eta_\nu/B_\nu \,\,{\rm or}\,\, \kappa^*_a = \eta^{\prime}_\nu/B_\nu\, ,
\label{kirchh}
\eeq
where $\eta^{\prime}_\nu$ is not corrected for final--state neutrino blocking. 
Furthermore, the net emissivity can be written as the sum of its {\it spontaneous} and {\it
induced} components:
\beq
\eta_\nu=\kappa_a\left[\frac{B_\nu}{1\pm\f\pr_\nu}+\left(1-\frac{1}{1\pm\f\pr_\nu}\right)I_\nu\right]\,\,  ,
\label{emission}
\eeq
where $+$ or $-$ is used for bosons or fermions, respectively. 

\section{Neutrino Cross Sections\label{cross6}}

Neutrino--matter cross sections, both for scattering and for absorption,  
play the central role in neutrino transport.  The major processes are
the super--allowed charged--current absorptions of $\nu_e$ and $\bar{\nu}_e$ 
neutrinos on free nucleons, neutral--current scattering off of free nucleons
(Schinder 1990; Janka \etal 1996; Burrows and Sawyer 1998ab; Reddy \etal 1998; Yamada 1998),
alpha particles, and nuclei (Freedman 1974; Leinson \etal 1988; Horowitz 1997; Burrows \etal 1981), 
neutrino--electron/positron scattering (Schinder and Shapiro 1982ab,1983), 
neutrino--nucleus absorption, neutrino--neutrino scattering, neutrino--antineutrino absorption (Janka 1991), and the inverses
of various neutrino production processes such as nucleon--nucleon bremsstrahlung
and the modified URCA process ($\nu_e + n + n \rightarrow e^- + p + n$).  
Compared with photon--matter interactions, neutrino--matter interactions
are relatively simple functions of incident neutrino energy.  
Resonances play little or no role and continuum processes dominate. Nice summaries
of the various neutrino cross sections of relevance in supernova 
theory are given in Tubbs and Schramm (1975) and in Bruenn (1985).
In particular, Bruenn (1985) discusses in detail  
neutrino--electron scattering and neutrino--antineutrino processes (see also Dicus 1972)
using the full energy redistribution formalism.  He also provides a serviceable approximation
to the neutrino--nucleus absorption cross section (Fuller 1982; Fuller, Fowler, 
and Newman 1983; Aufderheide \etal 1994). 
Recall that for a neutrino energy of $\sim$10 MeV the ratio of the charged--current cross section
to the $\nu_e$--electron scattering cross section is $\sim$100.
However, neutrino--electron scattering does play a role, along with neutrino--nucleon
scattering and nucleon--nucleon bremsstrahlung, in the energy equilibration of emergent $\nu_\mu$ neutrinos, 
though the relative contribution of each has yet to be determined. 
In this context, our current lack of an energy redistribution algorithm should be borne
in mind.  Nevertheless, our general conclusions in \S\ref{numu} concerning the $\nu_\mu$ neutrinos, their
softer than previously-believed spectra, the likely role
of bremsstrahlung in their production, and the consequences of their high scattering albedos, will only   
be strengthened when competent energy redistribution is included.  

\subsection{Charged-Current Absorption\label{CCabs}}  
The cross section per baryon for either $\nu_e$ or $\bar{\nu}_e$ absorption 
on free nucleons is larger than that for any other process. 
Given the large abundances of free neutrons and protons in protoneutron star atmospheres,
these processes are central to $\nu_e$ neutrino transport. 
A convenient reference neutrino cross section is $\sigma_o$, given by
\beq
\sigma_o\,=\,\frac{4G^2(m_ec^2)^2}{\pi(\hbar c)^4}\simeq \,1.705\times
10^{-44}\,cm^2\,\, .
\eeq
The total $\enu\,+\,n\, \righta\, e^-\,+\,p$ absorption cross section is then given by
\beq
\sigma^a_{\enu
n}=\,\sig_o\left(\frac{1+3g_A^2}{4}\right)\,\left(\frac{\varepsilon_{\nu_e}+\Delta_{np}}{m_ec^2}\right)^2\,
\Bigl[1-\left(\frac{m_ec^2}{\varepsilon_{\nu_e}+\Delta_{np}}\right)^2\Bigr]^{1/2}W_M\,\, .
\label{ncapture}
\eeq

The corresponding absorption cross section for the process, $\aenu\,+\,p\, \righta\, e^+\,+\,n$, is
\beq
\sigma^a_{\aenu
p}=\sig_o\left(\frac{1+3g_A^2}{4}\right)\,\left(\frac{\vep_{\bar{\nu}_e}-\Delta_{np}}{m_ec^2}\right)^2\,
\Bigl[1-\left(\frac{m_ec^2}{\vep_{\bar{\nu}_e}-\Delta_{np}}\right)^2\Bigr]^{1/2}W_{\bar{M}}\, .
\label{pcapture}
\eeq
$g_A$ is the axial--vector coupling constant ($\sim -1.26$) and $\Delta_{np}=m_nc^2-m_pc^2=1.29332$
MeV.  $W_M$ is the correction for weak magnetism and recoil (Vogel 1984),
never before included in supernova simulations, and is approximately equal to $(1 + 1.1\varepsilon_{\nu_e}/m_nc^2)$   
for $\nu_e$ absorption on neutrons.  At $\varepsilon_{\nu_e} = 20$ MeV, this correction is only $\sim2.5$\%.  
The corresponding correction ($W_{\bar{M}}$) for $\bar{\nu}_e$ neutrino absorption
on protons is $(1 - 7.1\varepsilon_{\bar{\nu}_e}/m_nc^2)$, which at 20 MeV is a large $-15$\%.   
To calculate $\kappa_a^*$, $\sigma^a_{\enu n}$ and $\sigma^a_{\aenu p}$ must be multiplied by the appropriate stimulated absorption 
correction, $1/(1-\f_{\nu_e}\pr)$ or  $1/(1-\f_{\bar{\nu}_e}\pr)$. Furthermore, final--state blocking
by either electrons or positrons and either protons or neutrons (\`a la eq. \ref{kirch03}) must be included.  
The consequences of these various terms for the neutrino spectra and neutrino-matter heating rates 
are explored in \S\ref{ntr}. Note that the sign of $\mu_e -\hat{\mu}$ 
in the stimulated absorption correction for $\bar{\nu}_e$ neutrinos         
is flipped, as is the sign of $\mu_e$ in the positron blocking term.  
Note also that the $\aenu\,+\,p\, \righta\, e^+\,+\,n$ process
dominates the supernova neutrino signal in proton--rich underground neutrino
telescopes on Earth, such as Super Kamiokande, LVD, and MACRO, a fact that emphasizes the
interesting complementarities between emission at the supernova and detection in \v Cerenkov and scintillation
facilities.

\section{Nucleon--Nucleon Bremsstrahlung \label{bremsst}}

A production process for neutrino/anti-neutrino pairs that has received but
little attention to date in the supernova context is neutral-current 
nucleon--nucleon bremsstrahlung ($n_1 + n_2 \righta n_3 + n_4 + \nu\bar{\nu}$).  
Its importance in the cooling of old neutron stars, for which the nucleons are quite
degenerate, has been recognized for years (Flowers, Sutherland, and Bond 1975), but only
in the last few years has it been studied for its potential importance in the atmospheres of 
protoneutron stars and supernovae (Suzuki 1993; Burrows 1997; Hannestad and Raffelt 1998). 
As a consequence, it has never before been incorporated into supernova codes. Neutron--neutron, proton--proton, and neutron--proton
bremsstrahlung are all important, with the latter the most important for symmetric matter.  As a source of $\nu_e$ and
$\bar{\nu}_e$ neutrinos, nucleon--nucleon bremsstrahlung can not compete  
with the charged--current capture processes.
However, for a range of temperatures and densities realized in 
supernova cores, it may compete with $e^+e^-$ annihilation as a source
for $\nu_\mu$, $\bar{\nu}_\mu$, $\nu_\tau$, and $\bar{\nu}_\tau$ neutrinos (``$\nu_\mu$''s).
The major obstacles to obtaining accurate estimates of the emissivity of this process are our poor knowledge
of the nucleon--nucleon potential, of the degree of suitability of the Born Approximation, and
of the magnitude of many--body effects (Hannestad and Raffelt 1998; Raffelt and Seckel 1991; 
Brinkman and Turner 1988).  Since the nucleons in protoneutron star atmospheres are not degenerate,
we present here a calculation of the total and differential 
emissivities of this process in that limit and assume a 
one-pion exchange (OPE) potential model to calculate the 
nuclear matrix element.  To acknowledge ignorance,
we encourage that a fudge factor of order unity, but perhaps as low as 0.1,  be appended to the rate.
The formalism we employ has been heavily influenced by those of Brinkman and Turner (1988) and Hannestad and Raffelt (1998),
to which the reader is referred for details and further explanations.  

Our focus is on obtaining 
a useful single--neutrino final--state emission (source) spectrum, as well as a final--state pair energy spectrum
and the total emission rate.  For this, we start with Fermi's Golden Rule for the total rate per neutrino species:
\beqa
Q_{nb}=(2\pi)^4\int \Bigl[\prod_{i=1}^4 \frac{d^3\vec{p}_{i}}{(2\pi)^3}\Bigr]
\frac{d^3\vec{q}_\nu}{(2\pi)^3 2\omega_\nu}
\frac{d^3\vec{q}_{\bar{\nu}}}{(2\pi)^3 2\omega_{\bar{\nu}}}\,
\omega\, \sum_{s}{|{\cal{M}}|}^2 \delta^4({\bf P}) \f_1\f_2(1-\f_3)(1-\f_4) \, ,
\label{bremfermi}
\eeqa
where $\delta^4({\bf P})$ is four--momentum conservation delta function, 
$\omega$ is the energy of the final--state neutrino pair,
($\omega_\nu$,$\vec{q}_\nu$) and ($\omega_{\bar{\nu}}$,$\vec{q}_{\bar{\nu}}$) 
are the energy and momentum of the neutrino and anti--neutrino, respectively,
and $\vec{p}_{i}$ is the momentum of nucleon $i$.  Final--state
neutrino and anti--neutrino blocking have been dropped. 

The necessary ingredients for the integration of eq. (\ref{bremfermi})
are the matrix element for the interaction and a workable procedure for handling
the phase space terms, constrained by the conservation laws.   We follow Brinkmann
and Turner (1988) for both of these elements. In particular, we assume for the
$n + n \righta n + n + \nu\bar{\nu}$ process that the 
matrix element is:

\beqa
\sum_{s}{|{\cal{M}}|}^2 = \frac{64}{4} G^2(f/m_\pi)^4 g_A^2 \Bigl[ (\frac{k^2}{k^2+m_{\pi}^2})^2 + \dots \Bigr ] 
\frac{\omega_{\nu} \omega_{\bar{\nu}}}{\omega^2} 
&=&A\frac{\omega_{\nu} \omega_{\bar{\nu}}}{\omega^2} \, ,
\label{matrixbrem}
\eeqa 
where the $4$ in the denominator accounts for the spin average for identical nucleons, $G$ is the
weak coupling constant, $f$ ($\sim1.0$) is the pion--nucleon coupling constant, $g_A$ is the axial--vector
coupling constant, the term in brackets is from the OPE propagator plus exchange and cross terms, $k$ is the nucleon
momemtum transfer, and $m_\pi$ is the pion mass.   In eq. (\ref{matrixbrem}), we have dropped $\vec{q}_\nu\cdot\vec{k}$
terms from the weak part of the total matrix element.  To further simplify the calculation, we set the
``propagator'' term equal to a constant $\zeta$, a number of order unity, and absorb into
$\zeta$ all interaction ambiguities.  The constant $A$ in eq. (\ref{matrixbrem}) remains.

Inserting a $\int \delta(\omega - \omega_{\nu} - \omega_{\bar{\nu}})d\omega$ by the neutrino phase space terms 
times $\omega \omega_{\nu} \omega_{\bar{\nu}}/{\omega^2}$ and integrating over $\omega_{\bar{\nu}}$ yields:

\beq
\int \omega \frac{\omega_{\nu} \omega_{\bar{\nu}}}{\omega^2} \frac{d^3\vec{q}_\nu}{(2\pi)^3 
2\omega_\nu}\frac{d^3\vec{q}_{\bar{\nu}}}{(2\pi)^3 2\omega_{\bar{\nu}}}\righta\frac{1}{(2\pi)^4}
\int_{0}^{\infty} \int_{0}^{\omega} \frac{\omega_{\nu}^2 (\omega - \omega_{\nu})^2}{\omega} d\omega_{\nu} d\omega \,  ,
\label{deltaneut}
\eeq
where again $\omega$ equals ($\omega_{\nu} + \omega_{\bar{\nu}}$).  If we integrate
over $\omega_{\nu}$, we can derive the $\omega$ spectrum.  A further integration over $\omega$ 
will result in the total volumetric energy emission rate.  If we delay such an integration, after
the nucleon phase space sector has been reduced to a function of $\omega$ and if we 
multiply eq. (\ref{bremfermi}) and/or eq. (\ref{deltaneut}) by $\omega_{\nu}/\omega$,  an integration
over $\omega$ from $\omega_{\nu}$ to infinity will leave the emission spectrum for the single final--state
neutrino.  This is of central use in multi--energy group transport calculations and 
with this differential emissivity and Kirchhoff's Law (\S\ref{stimabs}) we can derive an absorptive opacity.

Whatever our final goal, we need to reduce the nucleon phase space integrals and to do this we use the
coordinates and approach of Brinkmann and Turner (1988).  We define new momenta: $p_+ = (p_1 + p_2)/2$, $p_- = (p_1 - p_2)/2$,
$p_{3c} = p_3 - p_+$, and $p_{4c} = p_4 - p_+$, where nucleons $1$ and $2$ are in the initial state.  Useful direction cosines 
are $\gamma_1 = p_+ \cdot p_-/|p_+||p_-|$ and $\gamma_c = p_+ \cdot p_{3c}/|p_+||p_{3c}|$. 
Defining $u_i = p_i^2/2mT$ and using energy and momentum conservation, we can show that:
\beqa
d^3p_1d^3p_2 &=& 8d^3p_+d^3p_- 
\nonumber \\
\omega &=& 2T(u_- - u_{3c})
\nonumber \\
u_{1,2} &=& u_+ + u_- \pm 2(u_+u_-)^{1/2}\gamma_1
\nonumber \\
u_{3,4} &=& u_{+} + u_{3c} \pm 2(u_+u_{3c})^{1/2}\gamma_c \, .
\label{upm}
\eeqa

In the non--degenerate limit, the $\f_1\f_2(1-\f_3)(1-\f_4)$ term reduces to $e^{2y} e^{-2(u_+ + u_-)}$,
where $y$ is the nucleon degeneracy factor.  Using eq. (\ref{upm}), we see that the quantity $(u_+ + u_-)$ is independent
of both $\gamma_1$ and $\gamma_c$.  This 
is a great simplification and makes the angle integrations trivial.
Annihilating $d^3p_4$ with the momentum delta function in eq. (\ref{bremfermi}), noting that $p_i^2dp = \frac{(2mT)^{3/2}}{2}u_i^{1/2}du_i$,
pairing the remaining energy delta function with $u_-$, and integrating $u_+$ from $0$ to $\infty$, we obtain: 
\beq
d Q_{nb} = \frac{Am^{4.5}}{2^8\times3\times5 \pi^{8.5}} T^{7.5} e^{2y} e^{-\omega/T} 
(\omega/T)^4 \Bigl[ \int_0^{\infty} e^{-x}(x^2 + x\omega/T)^{1/2} dx\Bigr] d\omega \, .
\label{ezz4}
\eeq
The variable $x$ over which we are integrating in eq. (\ref{ezz4}) is equal to $2u_{3c}$.  That integral is analytic and
yields:
\beq 
\int_0^{\infty} e^{-x}(x^2 + x\omega/T)^{1/2} dx = \eta e^{\eta}K_1(\eta)\, ,
\label{kintegral}
\eeq
where $K_1$ is the standard modified Bessel function of imaginary argument, related to the Hankel functions, and
$\eta = \omega/2T$.  Hence, the $\omega$ spectrum is given by:
\beq
\frac{d Q_{nb}}{d\omega} \propto e^{-\omega/2T} \omega^5 K_1(\omega/2T) \, .
\label{omegaspect}
\eeq
It can easily be shown that $\langle \omega \rangle = 4.364 T$ (Raffelt and Seckel 1991).
Integrating eq. (\ref{ezz4}) over $\omega$ and using the thermodynamic identity in the non--degenerate limit:
\beq
e^y = \Bigl(\frac{2\pi}{mT}\Bigr)^{3/2} n_n/2 \, ,
\eeq
where $n_n$ is the density of neutrons (in this case), we derive for the 
total neutron--neutron bremsstrahlung emissivity of a single neutrino pair:
\beq
Q_{nb} = 1.04\times10^{30} 
\zeta(X_n \rho_{14})^2 (\frac{T}{{\rm MeV}})^{5.5} \, {\rm ergs\, cm^{-3}\, s^{-1}} \, ,
\label{bremssr}
\eeq
where $\rho_{14}$ is the mass density in units of $10^{14}$ gm cm$^{-3}$ and  
$X_n$ is the neutron mass fraction.  Interestingly,  this is 
within 30\% of the result in Suzuki (1993), even though he has substituted, without much justification, $(1+\omega/2T)$ for
the integral in eq. (\ref{ezz4}). ($[1+(\pi\eta/2)^{1/2}]$ is a better
approximation.)  The proton--proton and neutron--proton processes can be handled similarly and the total
bremsstrahlung rate is then obtained by substituting $X_n^2 + X_p^2 + \frac{28}{3} X_n X_p$ for
$X_n^2$ in eq. (\ref{bremssr}) (Brinkmann and Turner 1988).  
At $X_n = 0.7$, $X_p = 0.3$, $\rho = 10^{12}$ gm cm$^{-3}$, and T = 10 MeV, and taking the
ratio of augmented eq. (\ref{bremssr}) to the total rate for $e^+e^-$ 
production of $\nu_{\mu}\bar{\nu}_{\mu}$ pairs (Dicus 1972),  
we obtain the promising ratio of $\sim 5\zeta$. 
Setting the correction factor $\zeta$ equal to $\sim0.5$ (Hannestad and Raffelt 1998), we find 
that near and just deeper than the $\nu_\mu$ neutrinosphere, the bremsstrahlung rate is larger 
than that for classical pair production.

If in eq. (\ref{deltaneut}) we do not integrate over $\omega_\nu$, but at the 
end of the calculation we integrate over $\omega$ from $\omega_\nu$ to $\infty$,
after some manipulation we obtain the single neutrino emissivity spectrum:
\beqa
\frac{d Q_{nb}^{\prime}}{d\omega_{\nu}} =  
2C \Bigl(\frac{Q_{nb}}{T^4}\Bigr) 
\omega_{\nu}^3 \int^{\infty}_{\eta_\nu}  \frac{e^{-\eta}}{\eta} K_1(\eta) (\eta - {\eta^b_\nu})^2 d\eta 
= 2C \Bigl(\frac{Q_{nb}}{T^4}\Bigr) 
\omega_{\nu}^3 \int^{\infty}_{1} \frac{e^{-2\eta^b_{\nu}\xi}}{\xi^3} (\xi^2-\xi)^{1/2} d\xi \, ,
\label{spectrum}
\eeqa
where $\eta^b_{\nu} = \omega_\nu/2T$, $C$ is the normalization constant equal 
to $\frac{3\times5\times7\times11}{2^{11}}$ ($\cong 0.564$), and for the second expression we have used the
integral representation of $K_1(\eta)$ and reversed the order of integration.  In eq. (\ref{spectrum}),
$Q_{nb}$ is the emissivity for the pair.

Eq. (\ref{spectrum}) is the approximate neutrino emission spectrum  due to nucleon--nucleon bremsstrahlung.
A useful fit to eq. (\ref{spectrum}), good to better than 3\% over the full range of important values of $\eta_{\nu}$, is:
\beq
\frac{d Q_{nb}^{\prime}}{d\omega_{\nu}} \cong 
\frac{0.234 Q_{nb}}{T} \Bigl(\frac{\omega_\nu}{T}\Bigr)^{2.4} e^{-1.1 \omega_{\nu}/T}\, .
\eeq
Setting $\zeta$ equal to 0.5, we have incorporated bremsstrahlung into our Feautrier transport
algorithm.   In \S\ref{numu}, we show how the emergent $\nu_{\mu}$ spectrum depends upon $\zeta$.

\section{Basic Neutrino Transport Results\label{ntr}}  

The formalism and microphysics described in \S\ref{sec2} through \S\ref{bremsst}
were used to calculate the neutrino radiation fields for two snapshot profiles
in temperature, density, electron fraction, and velocity.
One of these is from the work of Burrows, Hayes, and Fryxell (1995) and represents 
the wind phase that follows explosion (Model W).
The second profile (our Model BM) was kindly provided to us by Tony Mezzacappa and is from a
multi--group, flux--limited diffusion simulation by Bruenn (Messer {\it et al.} 1998), 106
milliseconds after the bounce of the core of the Weaver and Woosley (1995)
15 M$_{\odot}$ star.  Since Messer {\it et al.} (1998) have already published 
their results for this model, in order to facilitate comparison we highlight our results for Model BM.  
Note that our focus is on neutrino atmospheres and not on completely self--consistent profiles and 
their evolution.  Hence, differences between the equations of state and microphysics employed in two 
different dynamical calculations, in particular any differences between the $\hat{\mu}$s, 
will translate at a given epoch into differences in composition and thermal profiles.  
Post--processing one group's snapshots with the code of another can lead to
differences in the neutrino fields that are larger than the differences  
in their thermal profiles.  The $\nu_e$ and $\bar{\nu}_e$ neutrino luminosity
profiles and spectra are particularly sensitive to differences between the $\hat{\mu}$s
used.  To check this, after achieving a steady state we turned on the $Y_e$ coupling for about 
5 milliseconds. The upshot was that $Y_e$ changed very little, demonstrating that we were using
substantially the same $\hat{\mu}$s as Messer {\it et al.} (1998).

We concentrate on the generic features of the energy, angle, and spatial distributions 
of the various neutrino radiation fields.  We use 50 energy groups, concentrating them 
below 50 MeV, so that the emergent spectra are well--resolved.  The models  
have 120 spatial grid points out to a radius of about 300 kilometers and we interpolate in
the various original models to resolve important features, such as the neutrinospheres
and the shock wave (for Model BM).  Since we are using the tangent--ray method to
set up and determine the angular grid, we employ from 119 to a few angular groups. In the code, we can
establish an arbitrary number of ``core rays" in the interior to increase the angular
coverage at small radii, but we found that we did not need to exercise this option.

The temperature ($T$), density ($\rho$), and $Y_e$ profiles for the two models are shown in Figure
1.  Model BM is a pre--explosion protoneutron star in a stalled shock configuration, while
Model W is a snapshot of a post--explosion neutrino--driven wind that expands off of the 
protoneutron star after explosion.  In Model W from 
Burrows, Hayes, and Fryxell (1995), $Y_e$ asymptotes to a value determined by the
the competition between $\nu_e$ and $\bar{\nu}_e$ neutrino absorption, $e^-$ and 
$e^+$ capture on nucleons, and the speed of expansion.  This situation is similar to that found
in a gas--dynamic laser or freeze out in the early universe.  The actual asymptotic $Y_e$ 
and acceleration timescale will depend, in a self--consistent calculation, on the details
of the neutrino--matter coupling and radiation fields and will be the subject of a future paper.
Also shown on the lower panel of Figure 1 are the neutron, proton, and alpha particle
mass fractions that bear on the physics of wind acceleration and the viability
of this wind as a site for the r--process (Woosley and Hoffman 1992; Qian and Woosley 1996).

\subsection{Optical Depths and Scattering Albedos versus Radius and Energy\label{albed}}

The integrated depth versus radius or interior mass provides a measure of the
global context of any transport problem.  Figure 2 shows the depth versus radius
and neutrino energy for $\nu_e$ neutrinos with energies
from 5 to 30 MeV in Model BM.  This is not the Rosseland mean which, due to the much higher 
average neutrino energies in the deep interior, reaches a value greater than $10^5$ at the center.
The position of the shock wave is manifest.  Figure 2 demonstrates that the 
position of the neutrinosphere ($\tau \sim 2/3$) is a stiff function of neutrino energy.  For $\nu_e$ neutrinos
and the energies depicted in the figure, the radii of the neutrinospheres range from $\sim 50$ kilometers
to $\sim$130 kilometers, more than a factor of two.  For the $\bar{\nu}_e$ and $\nu_{\mu}$ neutrinos,
the range is similarly broad, though due to the weaker neutrino--matter coupling for these
neutrinos the radii are correspondingly smaller.  These facts emphasize the dubious merit
of referring to a single neutrinosphere for a given species.  Figure 3 depicts the positions of the
neutrinospheres versus energy and type.  In Model BM, while 10 MeV $\nu_e$ neutrinos
decouple at $\sim$80 kilometers, 100 MeV neutrinos decouple as far out as the position of the shock.
This situation has a bearing on the strength of the high--energy spectral tail.
Note that for the $\nu_e$ and $\bar{\nu}_e$ neutrinos the gain region for Model BM, between 
$\sim$110 kilometers and the shock, resides at optical depths below $\sim 0.1$ near the peak
of their respective emergent spectra.  For slightly higher neutrino
energies, the optical depth of this region is correspondingly higher.  
Hence, energy deposition in this semi--transparent region is problematic
and requires a full transport code to study adequately.

It is important to distinguish absorption from scattering.  The scattering albedo 
is the a priori probability that an interaction is a scattering ($\kappa_{\nu}/\chi_{\nu}$).
It is a function of composition, neutrino energy, neutrino type, and final--state blocking.  
For $\nu_e$ neutrinos, the excess of neutrons over protons in the free--nucleon, high--entropy region interior
to the shock results in an albedo near 0.25, while for the $\bar{\nu}_e$ neutrinos it is $0.5-0.6$.    
Figure 4 depicts the Model BM scattering albedos versus radius as a function of 
energy for $\nu_e$ and $\bar{\nu}_e$ neutrinos.  In the interior,
the absorption process, $\nu_e + n \rightarrow e^- + p$, is suppressed by blocking due to final--state electrons.
This results in an elevated scattering albedo for the lower energy $\nu_e$ neutrinos. 
For $\nu_{\mu}$ neutrinos in Model BM, scattering predominates and exterior to 20 
kilometers the albedo is above 0.95.  Such a scattering albedo for the $\nu_{\mu}$ neutrinos
makes its transport a thermalization depth problem that can not be easily handled with
flux limiters.

\subsection{Emergent Spectra, Luminosities, and Heating Rates\label{heat}}

The emergent neutrino spectra and luminosities are functions of progenitor and they evolve.
Generally, the spectra after bounce harden with time (Mayle, Wilson, and Schramm 1987), 
but after hundreds of milliseconds or as accretion reverses into explosion 
(or otherwise subsides), the spectra start to soften.
The residue then cools inexorably over many seconds, like a 
clinker plucked from a smelter (Burrows and Lattimer 1986).   
Our Models are merely snapshots, but they serve as contexts in which to study the 
influence of various effects and physics.  In addition, the results can serve
as benchmarks against which to compare those from approximate schemes (see \S\ref{limit}).  The luminosity profiles
and spectra for Model BM are depicted in Figures 5 and 6, respectively. 
The $\nu_{\mu}$ neutrino luminosity includes that due to $\nu_{\mu}$, $\bar{\nu}_{\mu}$,
$\nu_{\tau}$, and $\bar{\nu}_{\tau}$ neutrinos.  The steeper rise and plateau
of the $\nu_{\mu}$ luminosity is a consequence of the small scattering albedo and deeper
point of energy decoupling, even though the $\tau = 2/3$ surface is at larger radii. 
The peaks in the $\nu_e$ and $\bar{\nu}_e$ luminosities mark the inner radius of the gain 
region, which resides where the luminosity slope is negative.  The rapid variation
in $\nu_e$ luminosity at smaller radii is a consequence of the variation in the   
temperature slope in the original model, itself presumably a consequence of sparse zoning.

The asymptotic $\nu_e$ and $\bar{\nu}_e$ luminosities are $4.3\times 10^{51}$ ergs s$^{-1}$
and $3.1\times 10^{51}$ ergs s$^{-1}$, respectively, 13\% higher and 9\% lower than
the corresponding ``BOLTZTRAN" numbers from Messer {\it et al.} (1998).  The 
differences must stem from a combination of differences in our numerical
algorithms, in our spatial, angular, and energy zoning, and in our cross sections.
Our emergent spectra for Model BM are given in Figure 6.  The hardness hierarchy of
$\nu_e < \bar{\nu}_e < \nu_{\mu}$ is manifest, as is the dominance of $\nu_{\mu}$
neutrinos at high energies.  The $\nu_e$ and $\bar{\nu}_e$ spectra can be fit
by a Fermi--Dirac distribution with temperatures and $\eta$s of 2.22 MeV and 3.16 for the $\nu_e$
neutrinos and 2.80 MeV and 3.48 for the $\bar{\nu}_e$ neutrinos.  The best Fermi--Dirac
fit to the $\nu_{\mu}$ neutrino spectrum has a negative $\eta$, which might as well be $-\infty$.
Note that the emergent $\nu_{\mu}$ spectrum shown in Figure 6 was calculated with the
bremsstrahlung $\zeta$ set equal to $0.5$.  The dependence of the $\nu_{\mu}$ spectrum
on $\zeta$ will be explored in \S\ref{numu}.

The corresponding energy--integrated inverse flux factors ($\int J_{\nu}d\varepsilon/\int H_{\nu}d\varepsilon$)
for Model BM are plotted versus radius in Figure 7. Figure 8 depicts the unintegrated
$\nu_e$ inverse flux factors ($J_{\nu}/H_{\nu}$) at given radii versus neutrino energy.
Since neutrino--matter heating terms are proportional to $J_{\nu}$, the higher the inverse flux factor
the more efficiently a given energy flux (luminosity) heats the matter in the semi--transparent
gain region.  Different transport algorithms result in different inverse flux factors, so getting this
term right can be important to the viability of the neutrino--driven 
supernova mechanism (Mezzacappa {\it et al.} 1998) and to the 
acceleration and entropy of the post--explosion wind (Burrows 1998a,b).
In addition, the harder the spectrum, the stronger the neutrino--matter coupling, so the 
$\nu_e$ and $\bar{\nu}_e$ neutrino spectra versus radius around and exterior to the 
neutrinospheres have a direct bearing on the heating rate.  

Figure 9 portrays
the $H_{\nu}$ and $J_{\nu}$ spectra as the $\nu_e$ neutrinos decouple.  
As this figure shows, at large radii $H_{\nu}$
and $J_{\nu}$ are the same, but at depth $J_{\nu}$ is much larger than $H_{\nu}$.  The precise
values of $J_{\nu}$ as the neutrinos decouple determine the matter heating rate. 
The energy--integrated heating and cooling rates versus radius for Model BM for 
all neutrino species individually are given in Figure 10.  The positions of radiative equilibrium are indicated
with a large dot and the inner radius of the gain region for each neutrino is denoted by an $X$.  
Note that the gain region identified on Figure 10 coincides with the gain region determined from
the luminosity plot (Figure 5).  Also included
on Figure 10 are the heating rates due to $\nu_e-\bar{\nu}_e$ annihilation and to
$\nu_{\mu}-\bar{\nu}_{\mu}$ and $\nu_{\tau}-\bar{\nu}_{\tau}$ annihilation, done properly
with the appropriate angular factors (Janka 1991).  
Aside from being competitive in the irrelevant unshocked regime,
heating due to neutrino pair annihilation is meager, at best.  In addition, due 
to the fuzziness of the neutrinospheres, the heating rate per cm$^{-3}$ does not follow the $1/r^8$ law
that might have been appropriate for a sharp neutrinosphere.  The difference between
the heating and cooling curves, the ``net gain," for Model BM 
is given by a solid line in Figure 11, to be compared with Figure 8 of Messer
{\it et al.} (1998).  We obtain similar heating rates throughout most of the 
supernova atmosphere, but slightly greater rates between 110 and 130 kilometers.
This slight difference could be due to a combination of things, including 
different techniques, different cross sections, slightly different equations of state, 
or our better angular and energy resolution.

\subsection{Consequences of Various Physical Terms}

The neutrino radiation fields depend upon terms that incorporate various physical
effects.  It is conceptually useful to gauge these terms by their influence
on the emergent spectra and on the heating rate.  Examples of effects that 
may or may not be included in simpler schemes are the final--state electron blocking
term for the charged--current absorption process (\S\ref{CCabs}), stimulated absorption
corrections (\S\ref{stimabs}), weak magnetism and recoil (\S\ref{CCabs}), and the 
velocity advection, aberration, and Doppler shift terms in the transport equation
(eqs. \ref{transport1} and \ref{east01}). The net gain and the $\nu_e$ and
$\bar{\nu}_e$ neutrino spectra for Model BM, with and without the blocking, 
weak magnetism and recoil, or the stimulated absorption terms,
are depicted in Figures 11 and 12. The blocking correction 
to the emergent $\nu_e$ luminosity is $\sim 15$\% and that
due to stimulated absorption is $\sim -3.5$\%.  
The blocking and stimulated absorption
corrections to the emergent $\bar{\nu}_e$ neutrino luminosity are of
opposite sign and approximately equal to $-6.0$\% and 3.5\%, respectively.  Blocking and
stimulated absorption shift the emergent $\nu_e$ and $\bar{\nu}_e$ spectra in opposite directions 
in a given energy group by as much as $\sim$20\% and $\sim -8$\%, respectively, due to blocking and $\sim -5.5$\% 
and $\sim 6.5$\%, respectively, due to stimulated absorption.
Blocking increases the net gain by 10--20\%, while stimulated absorption
decreases it by less than 5\%. Without electron blocking, the
absorption cross sections are artifically enhanced.  Since the degree of degeneracy
is different in the envelope and around the neutrinospheres, the magnitude of this effect
in the two regions is different, enhancing the emergent luminosity more than   
it decreases the coupling in the periphery.  Stimulated absorption has the opposite
effect (\S\ref{stimabs}), but its effect in the core and in the envelope is similarly
differential. 

In these Model BM calculations, the effects of weak magnetism and recoil on the 
emergent $\nu_e$ and $\bar{\nu}_e$ neutrino spectra and luminosities are small ($\le 2.0$\%).
This is due in part to the fact that 
the presence of scattering mutes the effect of changes in the absorption
cross section through the thermalization depth effect.  Due to the modest scattering albedo
(Figure 4), the response of the radial dependence of the radiation field to changes in
the absorption cross section is not linear with the change in the absorption 
cross section itself.  The increase in the net gain that one would anticipate due to any increase in the
$\bar{\nu}_e$ luminosity is countered by the concommitant decrease in        
the absorption cross section in the gain region. 

The winds that emerge from protoneutron stars after their envelopes supernova
are powered by neutrino energy deposition in the expanding gas.  Just as with
the supernova itself, the wind mass and enthalpy fluxes, velocities, entropies,
and compositions are influenced by details of neutrino--matter coupling
and neutrino transport.  The distribution of the heating determines
the magnitude, spatial extent, and timescale of acceleration.  
In turn, the degree of r--processing in the ejecta
is a function of the expansion timescale, the asymptotic $Y_e$, and
entropy (Woosley and Hoffman 1992; Qian and Woosley 1996).
Hence, it is important to gauge the relative strengths of the various terms
that determine the degree and distribution of neutrino--matter heating.

The effect of the velocity terms on the emergent $\nu_e$ and $\bar{\nu}_e$ spectra 
for Model W is depicted in Figure 13.  Model W is the post--explosion
wind model from Burrows, Hayes, and Fryxell (1995), in which the 
velocities at large radii are $\sim$30,000$\,{\rm km \, s^{-1}}$. Of course, at small 
radii they are zero.  As Figure 13 shows, the velocity effects collectively boost 
the emergent spectra of Model W in a given energy group by $\sim$15\%, with a corresponding boost
in the $\nu_e$ and $\bar{\nu}_e$ luminosities by 15\% and 13\%, respectively.
This is mostly a consequence of the Doppler shift of the radiation field.
Due to the smaller 
velocities in the important accretion regions in Model BM, the velocity
corrections for that model are much smaller ($\le 5$\%).  Figure 14 shows the net 
gain in Model W for our fiducial model, as well as without blocking, 
velocity corrections, or weak magnetism/recoil corrections and implies 
that various terms not easily or often included in flux--limited or energy--integrated
transport can each make a $\sim 10$\% difference in the parameters of the wind.
Note that though the weak magnetism/recoil corrections for Model BM are
small, those for Model W are modest.  
This result implies that the importance of absorption cross section 
changes is a function of the specific thermal and composition profiles.
What distinguishes the wind is the more abrupt transition from the diffusive to the streaming
regime and the lower $Y_e$ value in its decoupling region (Figure 1).
Whereas, in Model BM the slight increase in the core luminosity due to
the inclusion of the weak magnetism/recoil term is nullified by the decrease
in the absorption opacity in the envelope when determining the net gain, in Model W
the slight increase in the luminosity due to the lower absorption cross section
that is a consequence of weak magnetism (particularly for the 
$\bar{\nu}_e$ neutrinos) is not adequate to counter the resulting
greater transparency of the envelope.  The $Y_e$ at the base of the 
wind's atmosphere is smaller than that near the neutrinospheres in Model BM, with
the result that the scattering albedo for the $\bar{\nu}_e$ neutrinos is larger
there.  This results in a smaller increase in the emergent luminosity
that can't compensate for the increase in the transparency of the wind's mantle.

The anisotropy of neutrino--nucleon scattering and the difference between the transport
and the total cross sections (eq. \ref{transcs})
can in principle translate into larger inverse flux factors and, hence,
greater net gain.  Backscattering increases $J_{\nu}$ for a given $H_{\nu}$ and delays
the transition from isotropic to forward--peaked radiation fields.  However, 
since absorption plays an important role for the $\nu_e$ and $\bar{\nu}_e$ neutrinos  
and their scattering albedos are not very close to one, the backscatter effect is muted.  The upshot 
is that anisotropy accounts for only $\sim 2$\% of the net gain and results in  
shifts of less than 1\% in the $\nu_e$ or $\bar{\nu}_e$ spectra.

\section{Flux Limiters\label{limit}}

It is common to seek methods for solving eqs. (\ref{east02}) and (\ref{east03})
without employing the full machinery of transport.  In principle, such methods
simplify the mathematics and speed solution, but they compromise accuracy.  Eqs. (\ref{east02}) and (\ref{east03})
are the first two equations in a moment hierarchy in which each moment equation involves
still higher-order moments; to solve such a hierarchy precisely requires the solution of an infinite
number of moment equations.  The simplifying ansatz often introduced is that higher-order moments 
can be written in terms of lower-order moments, thereby closing the 
system of equations.  However, these so-called closure relations can vary a great deal. 
The most common closure is the flux-limiter.

In flux-limited schemes, eq. (\ref{east03})
is reduced to its diffusion form in which $H_{\nu}$ is set equal to the product of the
gradient of $P_{\nu}$ and a coefficient (the flux limiter),  $P_{\nu}$ is set equal to $1/3 J_{\nu}$ (the Eddington closure),
and this expression for $H_{\nu}$ is inserted into the divergence term in eq. (\ref{east02}).
Only this zeroth-moment (energy) equation is solved. 
The third-order moment, $N_{\nu}$, important for multi-energy group calculations in moving media,
is generally ignored.  The art of this approach is in the choice of the flux limiter, so-called
because the coefficient is constructed in such a way that the flux, otherwise mathematically diffusive in this scheme, 
does not exceed the streaming limit, $J_{\nu}$.  This is necessitated by the 
dropping of eq. (\ref{east03}), with its causality-enforcing time derivative.  
All angular information is lost, though some flux limiters are derived under certain
assumptions about the angular distribution ({\it cf.} Levermore and Pomraning 1981).   
Examples of flux-limiters that have been employed in supernova calculations are
Wilson's (Bowers and Wilson 1982) and Bruenn's (1985; Messer \etal 1998).  Their prescription
for the flux is:
\beq
\hnu = -\frac{\lambda_\nu}{3} \left(1 + \frac{|R|}{3} \Xi_{F}\right)^{-1} \frac{\p J_{\nu}}{\p r}\, ,
\label{wlimit}
\eeq
where $\Xi_{F} = 1 + 3/(1 + |R|/2 + R^2/8)$ for Wilson and $\Xi_F = 1.0$ for Bruenn, $R = -\lambda_{\nu} \frac{\p ln J_{\nu}}{\p r}$,
and $\lambda_{\nu}$ is the transport mean-free-path.  The expression in parentheses 
is the limiter.  Hence, in flux-limiter closures, $H_{\nu}$ is a function of only $J_{\nu}$ and
its derivative (or $J_{\nu}$ and $R$) and $p_{\nu}$ ($=\pnu/\jnu$) is set equal to $1/3$.  
Note that the Knudsen parameter ($R$) is small in the diffusion limit.
The subscript $\nu$ is a reminder that these equations apply for each energy group.
It is important to note what should be obvious: not all flux limiters  
are the same, nor are their quantitative consequences.  Hence, there is 
not a uniform ``flux limiter" result.  For instance, the MGFLD scheme of Messer \etal (1998)
is merely one such approach.  The results of employing a given flux limiter
can deviate from the correct result as much as can the results derived using various flux limiters  
deviate from one another. 

\subsection{Angular Distributions}

To gauge the character of the proper angular dependence of the neutrino distribution functions 
in the supernova context, we provide in Figure 15 the Model BM Eddington
factor ($p_{\nu}$) versus radius for $\nu_e$ neutrinos, 
at particle energies from 5 to 30 MeV.
As we would expect from the decoupling hierarchy, the $\nu_{\mu}$ Eddington factors
start their rise from the isotropic value of $1/3$ from the deepest layers.
However, for all neutrino species, particularly for the $\bar{\nu}_e$ and $\nu_{\mu}$
neutrinos, the Eddington factor is a stiff function of energy and only gradually makes the transition
from $1/3$ to $0.75$ over a region that can be 50 to 150 kilometers wide.
Many flux limiters effect the related transition from diffusion to free streaming early (at higher $\tau_\nu$s) and 
within an unphysically narrow range in radius and $\tau_\nu$.  This can be seen in Figure 16 where 
the Bruenn and Wilson flux limiters for $\nu_e$ neutrinos at 20 MeV in Model BM are compared with the ``effective" 
flux limiter derived using full transport.  Approximately 20 kilometers interior to the appropriate point, 
the Bruenn and Wilson limiters begin to deviate from the diffusion value of 1.0. 

%
%

Polar plots depicting representative angular distributions of the Model BM $\bar{\nu}_e$ specific
intensity ($I_{\nu}$) field for an energy of 15 MeV are presented as solid lines in Figure 17.
The transition from isotropy to forward--peaked is clear, as is the gradual
nature of that transition.  There is no 
corresponding angular function for either the Bruenn or Wilson limiter.

Complementary to this polar plot are
Figures 18 through 20 of the full transport phase--space densities ($\f_{\nu}$) versus energy
at various radii and for all the neutrino species.  Depicted are the
phase--space densities along the 0$^{\circ}$ and 90$^{\circ}$ directions.
For the $\nu_e$ neutrinos, the degree of degeneracy at depth is clear; one can
almost read the $\nu_e$ chemical potentials off the graph.  From Figure 19, we see
that the occupancy of the $\bar{\nu}_e$ neutrino states is generically low, but from Figure 20
we see that at depth and for low energies the occupancy of the $\nu_{\mu}$
neutrino states can approach 0.5.  Blocking due to final--state $\nu_{\mu}$ occupancy
is generally unimportant in the pair source terms, since the peak energies of the pair source functions
are always significantly above the energies at which $\f_{\nu}$ is high.  

Among other things, Figures 18 through 20 convey a sense of the angular 
dependence of $\f_{\nu}$, ignored in the standard flux-limiter schemes.
At depth, since the radiation fields are isotropic, the 0$^{\circ}$ and 90$^{\circ}$
curves are the same.  However, with increasing radius and at lower energies, deviations
from isotropy manifest themselves; transverse beams are less occupied than forward
beams.  As expected, at low optical depths this differential effect is quite pronounced.
Flux-limited transport schemes are not capable of addressing or illuminating this
phenomenology.

\subsection{Neutrino Heating and Emergent Spectra Using Flux Limiters}

Given the nuances in the angular distributions of the neutrino radiation fields
portrayed in Figures 15-20, it is no wonder that flux limiters only inadequately
represent the radiation energy density profiles, emergent spectra, and net gain in the
semi-transparent decoupling region.  This is made manifest by comparing 
the emergent spectra and net gain derived using such flux limiters with 
those same quantities obtained using full transport.
%
%
Figure 21 compares the emergent $\nu_e$ spectrum for the BM model using the 
full Feautrier/tangent-ray formalism, Bruenn's limiter, and Wilson's limiter.   
Though comparisons for many snapshot profiles would be useful, we
can still conclude for such an early protoneutron star epoch that Bruenn's limiter, 
as simple as it is, results in a spectrum that deviates from the more precise spectrum
by $\sim$5-10\%, while Wilson's limiter, despite its modestly greater
complexity, can be off by as much as $\sim$20-30\%.  Figure 22, in which curves
of the net gain (heating) versus radius are compared, tells a similar story:
Bruenn's limiter yields net gains that are generally off, but by no more than $\sim$20\%, 
while Wilson's can be off by as much as $\sim$50\%.  Figures 21 and 22
serve to illustrate both that all flux limiters are not the same and, in particular, that they
can underestimate the net gain in the outer gain region by many tens of percent.   

In sum, flux-limiter schemes can miscalculate net heating rates, radiation energy densities 
(quantities that factor into the net gain), emergent spectra, and the inverse flux factors by from 5\% to 50\%
and can artificially accelerate the transition from isotropy to free--streaming
in the $\tau < 1$ region.  This is particularly true for neutrinos, with their extended
neutrinospheres.  Furthermore, the thermalization depth effect is difficult to handle with flux
limiters when the scattering albedo is large.  The albedo for $\nu_{\mu}$ neutrinos
is above 0.90 throughout most of the object.  As a result, only full transport
can properly handle the enhancement in the effective absorption path due to the frustrated
escape caused by scattering.

\section{Determinants of the Emergent $\nu_{\mu}$ Neutrino Spectrum\label{numu}}

The $\nu_{\mu}$ and $\nu_{\tau}$ neutrinos and their antiparticles carry
away from the protoneutron star more than 50\% of its total binding energy.
Since they do not participate in charged--current interactions, they energetically decouple
at smaller radii and, hence, at larger temperatures, than the other
neutrino species.  This results in a harder spectrum (Figure 6) and the 
hardness hierarchy alluded to in \S\ref{heat}.  The fact that there are four 
species is primarily responsible for their major cooling role.  Neutrino--matter
energy coupling is affected by the inverse production processes 
of pair annihilation and nucleon--nucleon
bremsstrahlung (\S\ref{bremsst}), as well as by neutrino--nucleon
and neutrino--electron scattering.  The proper treatment of energy redistribution
by scattering is deferred to a later publication.  However, it is clear that 
scattering generically softens the $\nu_{\mu}$ spectra. 
  
Ignoring the potential effects of neutrino oscillations,
the emergent $\nu_{\mu}$ spectra have a direct bearing on the process of neutrino
nucleosynthesis (Woosley {\it et al.} 1990) and on the suitability of various 
underground detectors that rely on neutral--current spallation processes with high
energy thresholds.  In both cases, the relevant neutral--current interaction cross sections  
are stiffly increasing functions of neutrino energy, with thresholds above $\sim 15$ MeV (Haxton 1990). 
Hence, they are most sensitive to the $\nu_{\mu}$ component and its precise spectrum.
In the past, people had thought that the $\nu_{\mu}$ spectra were hard, with effective
Fermi--Dirac temperatures of $\sim 8-9$ MeV and average energies of $\sim 25-30$ MeV.
However, the $\nu_{\mu}$ energy spectrum on Figure 6 can be very 
approximately fit with a temperature of 7 MeV.  

Bremsstrahlung has a major effect on the $\nu_{\mu}$ radiation field.
The factor $\zeta$ in \S\ref{bremsst} incorporates
a correction for our approximations to the propagator terms and 
to the nuclear matrix element.   In Figure 6, $\zeta$ was set equal to $0.5$.
Using Hannestadt and Raffelt (1998) and our own estimates of the correct propagator terms,
we derive that above $10^{13}$ g cm$^{-3}$ $\zeta$ is above 0.7 and that at and
around $10^{11}$ g cm$^{-3}$ $\zeta$ is near 0.2.  This translates into an ``average"
$\zeta$ of $\sim 0.5$ for protoneutron stars.  Figure 23 depicts the consequences of varying
$\zeta$ from 0.0 to 1.0 in steps of 0.2 for the emergent $\nu_{\mu}$ energy spectrum.  
Due to the presence of an absorption term for every emission term (Kirchhoff's Law; eq. \ref{kirchh}),
the strength of the spectrum is not strictly linear in $\zeta$.
As Figure 23 demonstrates, nucleon--nucleon bremsstrahlung is softer than $e^+e^-$ annihilation
(the other major $\nu_{\mu}\bar{\nu}_{\mu}$ source) and can dominate at low energies.
Though the emergent spectra are softer, due to the extra source the spectra are also brighter at every energy.
Hence, the inclusion of nucleon--nucleon bremsstrahlung increases the flux,
while decreasing the average and peak neutrino energies. This is important.
At 10 MeV, the $\nu_{\mu}$ spectrum can be more than a factor of {\it two} stronger
with bremsstrahlung than without.  For energies above $\sim 35$ MeV, $e^+e^-$
annihilation still dominates the emergent spectrum.
In Figure 23, the lowest curve corresponds to a pure
$e^+e^-$ annihilation source.  Note that it is demonstrably harder than when 
$\zeta$ is large and that it alone is ``pinched."
Though it still remains to be determined whether nucleon--nucleon 
bremsstrahlung in supernovae is in fact dominant for $\nu_{\mu}$ spectrum formation,
Figure 23 suggests that it is, particularly at lower neutrino energies.
Since energy transfer due to neutrino--matter scattering and gravitational
redshifts will only further soften the emergent spectra, we conclude that $\nu_{\mu}$
spectra are indeed softer than traditionally quoted.

Also shown on Figure 23 is an emergent $\nu_{\mu}$ spectrum with the scattering cross 
sections very artificially cut by one half.  This curve demonstrates the severe dependence of
the spectra on the basic numbers associated with the neutrino--matter interaction.
It suggests that if we did not have a fairly good handle on the basic interactions
of neutrinos with nucleons our predictions would be quite different, and perhaps 
would be quite wrong.

\section{Summary}

We have constructed and described an implicit, multi--group, multi--angle,
multi--species neutrino transfer code to be used in the context of core--collapse supernovae 
and protoneutron stars. The basic algorithm embodies the Feautrier and tangent--ray approaches to 
spherical atmospheres and is conceptually equivalent to various Boltzmann solvers.
It is capable of resolving angular distributions and of calculating angular 
moments to great precision and employs accelerated ${\bf \Lambda}$ iteration to achieve
rapid convergence.  Focusing on ``neutrino atmospheres,"  
we presented the energy spectra, neutrino heating rates, Eddington factors, angular distributions, and phase 
space densities for typical post--bounce structures.  The influence on these quantities, in particular
on the net gain, of various corrections to the charged-current cross section and 
terms in the transport equation were examined and the character 
of the neutrino radiation fields and spectra was scrutinized.  One goal has been to provide a detailed
snapshot of the neutrino radial, angular, energy, and species distributions in
a typical post--bounce environment, including in the protoneutron star context, and to explore the factors that determine the 
heating rates in the semi--transparent gain region, so central to the viability of the
neutrino--driven mechanism of supernova explosions.  To this end, we focused on the decoupling 
transition of the emergent neutrinos.  Moreover, we compared the emergent spectra and neutrino
heating rates obtained using representative flux limiters with those obtained using our Feautrier
transport algorithm to gauge the accuracy of those oft-used approximate schemes.
Finally, we derived the rate of nucleon--nucleon bremsstrahlung and its neutrino
source spectrum and showed for the first time that it probably dominates
$\nu_{\mu}$ neutrino production and spectrum formation.

The tool that we have developed is meant to explore supernova explosions, protoneutron star cooling,
the neutrino signature of core--collapse, neutrino shock break--out, and post--explosion winds,
among other things.  It is also easily converted into a photon transport code for the study
of classical supernova light curves.  However, we have yet to generalize the scheme for use in
multi--dimensional supernova simulations or in the general relativisitic context, 
nor have we parallelized it for use on shared--memory machines.  Hence, much technical work remains.

Supernova theory has been evolving for thirty years and in that time our understanding 
of the neutrino and its interactions has changed substantially.  There are now indications from 
atmospheric and solar neutrino experiments that lepton number is not strictly conserved
and that neutrinos may mix.  Heating in the protoneutron star 
mantle is a subtle sum of competing effects.  We have investigated in this paper
but a few of these.  This effort to fully characterize the neutrino radiation fields 
is part of a larger effort, as yet unfinished, to understand the mechanism
of supernova explosions and their systematics.  However, when this puzzle box is eventually opened,
precise neutrino transport will certainly be one of the keys.

\acknowledgements

We are pleased to acknowledge Ray Sawyer, Georg Raffelt, Dimitri Mihalas, 
Tony Mezzacappa, Chuck Horowitz, John Hayes, and Steve Bruenn for useful contributions
and guidance.  In addition, we extend special thanks to Jorge Horvath for his help 
with the bremsstrahlung calculations and to Steve Bruenn and Tony Mezzacappa for providing
some of their output in electronic form.  This work was supported by the NSF under
grant AST96-17494 and was performed under the auspices of the U.S. Department of Energy by
Lawrence Livermore National Laboratory under Contract W-7405-Eng-48.

\clearpage

\figcaption{The temperature ($T$), density ($\rho$), electron fraction ($Y_e$), 
proton fraction ($Y_p$), neutron fraction ($Y_n$), and alpha fraction ($Y_{\alpha}$) for Models 
BM (top panel) and W (bottom panel). In Model BM, the shock is located at
170 km, but in Model W there is no shock on the grid.
\label{fig:1}}

\figcaption{The $\nu_e$ neutrino optical depth ($\tau_{\nu_e}$) versus 
radius (in kilometers) for Model BM, at various particle energies.
As the energy of the neutrino increases the degree of transparency decreases.
The dip in the optical depth at 170 km is where the shock (and, hence, a density jump) is located.
The solid horizontal line shows where $\tau_{\nu_e}$ = 2/3.
\label{fig:2}}

\figcaption{The neutrinosphere radii versus neutrino energy for $\nu_e$, $\bar{\nu}_e$, and
``$\nu_\mu$'' neutrinos. For a given neutrino energy, the ``$\nu_\mu$'' neutrinos decouple
first, resulting in a $\tau=2/3$ radius that is smaller than that for either $\nu_e$ or $\bar{\nu}_e$
neutrinos.  The hierarchy in decoupling radii of $\nu_e > \bar{\nu}_e > \nu_{\mu}$ 
is manifest.
\label{fig:3}}

\figcaption{The $\nu_e$ and $\bar{\nu}_e$ scattering albedos versus 
radius for neutrino energies of 10, 20, and 40 MeV.
The shock is at $\sim$170 km.  The increases in the $\nu_e$ albedo at
small radii can be traced to $e^-$ blocking of $\nu_e$ absorption 
on neutrons, predominantly.
\label{fig:4}}

\figcaption{Model BM $\nu_e$, $\bar{\nu}_e$, and $\nu_\mu$ luminosities versus radius (in km).
The ``$\nu_{\mu}$'' luminosity is the sum of the $\nu_{\mu}$, $\bar{\nu}_{\mu}$, $\nu_{\tau}$, 
and $\bar{\nu}_{\tau}$ neutrino luminosities.  The modest peaks mark the inner radius of the gain region,
in which, due to net absorption, the luminosity slope is gently negative.
\label{fig:5}}

\figcaption{The Model BM emergent neutrino luminosity spectra for the three neutrino types.
The symbols indicate the positions of the energy groups ($\nu_e$ - filled squares;
$\bar{\nu}_e$ - filled triangles; $\nu_\mu$ - open circles).  
\label{fig:6}}

\figcaption{The energy--integrated inverse flux factor ($\langle 1/F_{\nu}\rangle$) as a 
function of radius for $\nu_e$ and $\bar{\nu}_e$ neutrinos.
The sharp increase in $\langle 1/F_{\rm VEF}\rangle$ at 110 km 
occurs just inside the gain radius, where the neutrinos
are starting to decouple from the matter. At large radii (off the plot), the 
inverse flux factors approach the unity expected for the free--streaming regime.
\label{fig:7}}

\figcaption{The Model BM ratios of the $\nu_e$ neutrino energy density to the $\nu_e$ 
energy flux for radii of 100, 125, 150, 180, 200, 220, and 300 km. 
\label{fig:8}}

\figcaption{The $\nu_e$ neutrino energy flux ($F_{\nu}$, solid) and energy density (c$E_\nu$, dashed)
spectra at various radii. A, B, C, D, and E denote radii at 50, 130, 200, 250, and 300 km. 
At depth, the spectra are very
different, but they converge at large distances from the neutrinospheres. At
these large distances, the unintegrated flux factor, $F_\nu$/c$E_\nu$, is unity.
\label{fig:9}}

\figcaption{Model BM heating and cooling rates (in ergs g$^{-1}$ s$^{-1}$) versus radius (in km). 
The heating and cooling rates for the three neutrino species are shown, along with the 
$\nu-\bar{\nu}$ annihilation energy deposition rates.   
The solid points indicate where radiative equilibrium is achieved for each 
neutrino species. The $X$'s indicate the positions of the gain radii 
for the respective neutrino types. The top two solid lines
are the heating and cooling curves for the $\nu_e$ neutrinos. The dashed lines are the 
heating and cooling curves for the $\bar{\nu}_e$ neutrinos. The bottom two solid
lines are the heating and cooling curves for $\nu_\mu$ neutrinos. The bold dashed
curves are the heating rates for the $\nu_e\bar{\nu}_e \rightarrow e^+e^-$ process (top)
and both the $\nu_{\mu}\bar{\nu}_{\mu} \rightarrow e^+e^-$ and 
$\nu_{\tau}\bar{\nu}_{\tau} \rightarrow e^+e^-$ processes (bottom).
\label{fig:10}}

\figcaption{The net heating rate (net gain) for various BM models versus radius. The fiducial model (solid) is compared
to models with no stimulated absorption (short-dashed line), with no e$^-$ blocking (long-dashed line), or
with no weak magnetism/recoil (dot-dashed line). The absence of either stimulated absorption or weak magnetism/recoil
would result in an increase in neutrino absorption and, thus, a greater heating rate.  The absence of $e^-$ blocking 
would result in a decrease in the net gain.
\label{fig:11}}

\figcaption{The emergent luminosity spectrum for both the $\nu_e$ and
$\bar{\nu}_e$ neutrinos for our fiducial Model BM (solid line),
compared with models without stimulated absorption (small--dashed line), e$^-$ blocking
(long--dashed line), or weak magnetism/recoil (dot--dashed line).
Also included is a model for which the total scattering cross section
is substituted for the transport cross section (short dash-long dash).
\label{fig:12}}

\figcaption{The emergent $\nu_e$ luminosity spectrum for the wind Model W with a velocity
field (solid line) and without a velocity field (dashed line), versus energy in MeV.
The velocity terms boost the spectrum that emerges from a protoneutron star with a wind
by $\sim$15\%.
\label{fig:13}}

\figcaption{The net heating rate (net gain) versus radius (in km) for wind Model W. The fiducial model (solid) is compared
with models for which either weak magnetism/recoil (short-dashed line) or e$^-$ blocking (long--dashed line)
is ignored.  Also included is a model for which the velocities were set equal to zero (dot--dashed).
\label{fig:14}}

\figcaption{The Eddington Factor for $\nu_e$ neutrinos versus radius (in km) at various $\nu_e$ neutrino energies. 
At depth, in the diffusive region the Eddington factors converge to 1/3.  At large radii, the Eddington 
factors approach unity. The low--energy neutrinos are the first to decouple
and their Eddington factors approach unity faster than those of the higher--energy neutrinos. 
\label{fig:15}}

\figcaption{Electron neutrino flux limiter profiles versus radius (in kilometers) for model BM 
using Bruenn's flux limiter (dot-dahsed), Wilson's flux limiter (solid), and an artifical flux
limiter derived from the full formalism (dashed).  The $\nu_e$ neutrino energy is 20 MeV.
See text and eq. (\ref{wlimit}) for details.
\label{fig:16}}

\figcaption{Polar plots of the specific intensity ($I_{\nu}$) of $\bar{\nu}_e$ neutrinos 
with an energy of 15 MeV. Shown are angular distributions at radii of 80, 120, 170,
and 300 km. In the interior, the radiation fields are isotropic and strong.  At large radii, 
the distribution becomes more forward--peaked and geometric dilution decreases $I_{\nu}$.
The numbers on the left axis provide the scale, with the negative numbers emphasizing
the fact that the rays are pointing backward in this hemisphere. 
\label{fig:17}}

\figcaption{The phase--space density ($\f_\nu$) for the $\nu_e$ neutrinos versus
neutrino energy in MeV, for various radii from 20 km to 170 km. The solid lines
are for the forward direction and the dashed lines are for the transverse direction
(at $\sim 90^{\circ}$ to the radial direction).  At small radii, the $\nu_e$ neutrinos are degenerate,
but at larger radii, and generally at larger energies, they quickly become non--degenerate.
Note that at small energies, ``larger'' radii, and large angles, the degeneracy of the 
$\nu_e$ neutrinos diminishes.
\label{fig:18}}

\figcaption{The phase--space density ($\f_{\nu}$) for the $\bar{\nu}_e$ neutrinos versus
neutrino energy in MeV, at radii of 120, 150, and 200 km. The solid lines
are for the forward direction and the dashed lines are for the transverse direction
(at $\sim 90^{\circ}$). 
\label{fig:19}}

\figcaption{The phase--space density ($\f_\nu$) for the $\nu_\mu$ neutrinos versus
neutrino energy in MeV, at radii of 55, 70, and 80 km. The solid lines
are for the forward direction and the dashed lines are for the transverse direction
(at $\sim 90^{\circ}$). At depth, and at lower energies, $\nu_\mu$ neutrino
degeneracy ($\f_\nu$) approaches 0.5, as expected for the situation with no net
$\nu_\mu$ lepton number and, hence, zero chemical potential.
\label{fig:20}}

\figcaption{The emergent $\nu_e$ neutrino luminosity spectrum for model BM using the full Feautrier/tangent-ray
formalism (solid), Bruenn's limiter (short dashed), and Wilson's limiter (long dashed) (cf. Figure 6).  
\label{fig:21}}

\figcaption{A comparison of the net gain (in erg g$^{-1}$ s$^{-1}$) versus radius (in kilometers) for model profile BM,
calculated using the full transport formalism of this paper (solid), Bruenn's flux limiter (dot-dashed),
and Wilson's flux limiter (dashed) (cf. Figure 11).
\label{fig:22}}

\figcaption{The emergent $\nu_{\mu}$ luminosity spectra for Model BM for bremsstrahlung factors,
$\zeta$, of 0.0, 0.2, 0.4, 0.6, 0.8, and 1.0.   Also included is the $\nu_{\mu}$ spectrum with 
the $\nu_{\mu}$--nucleon scattering cross section artificially decreased by 50\%. 
\label{fig:23}}

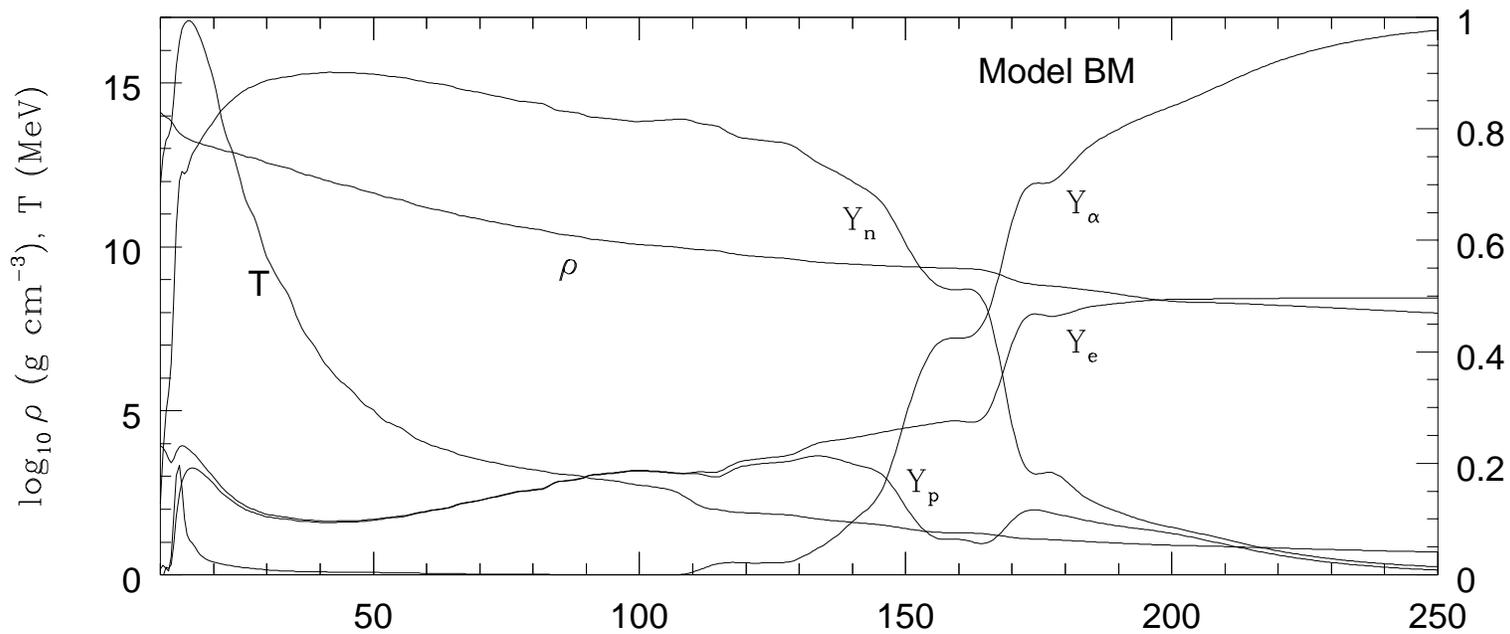
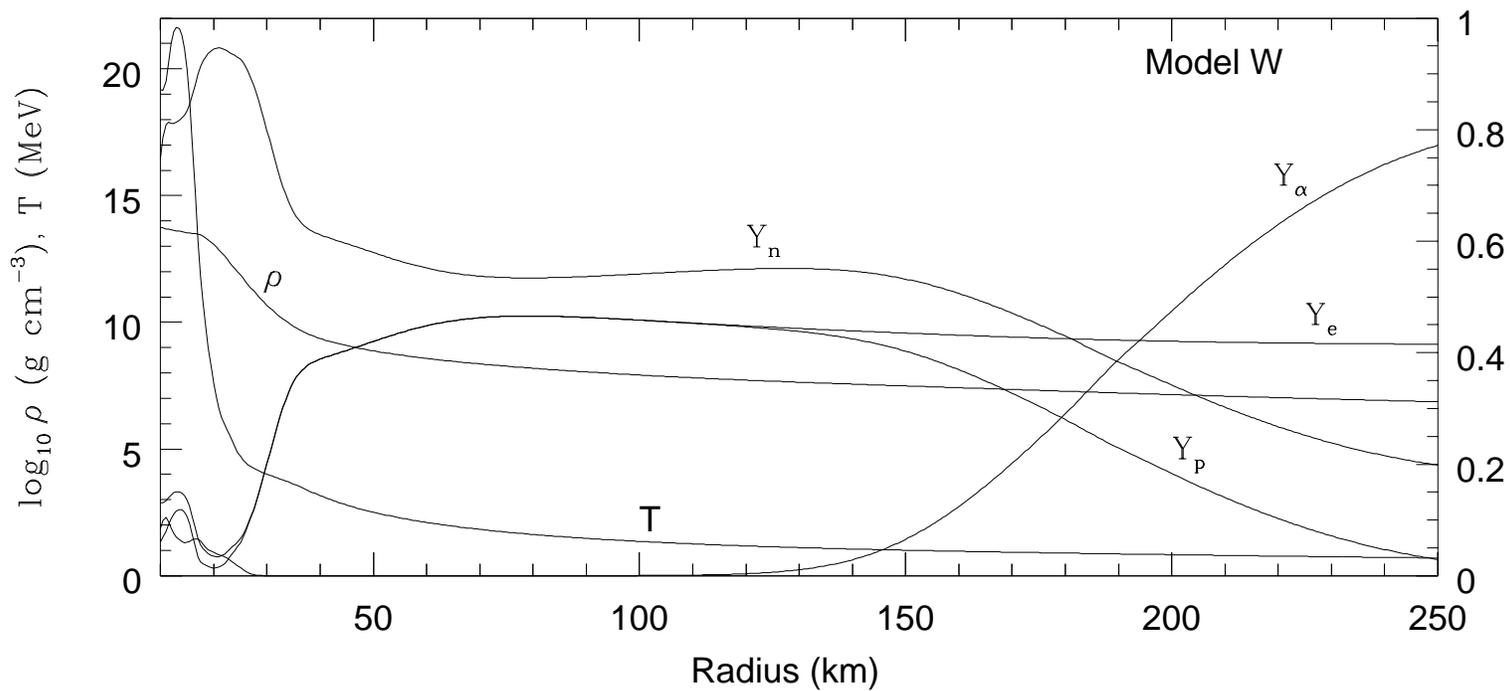
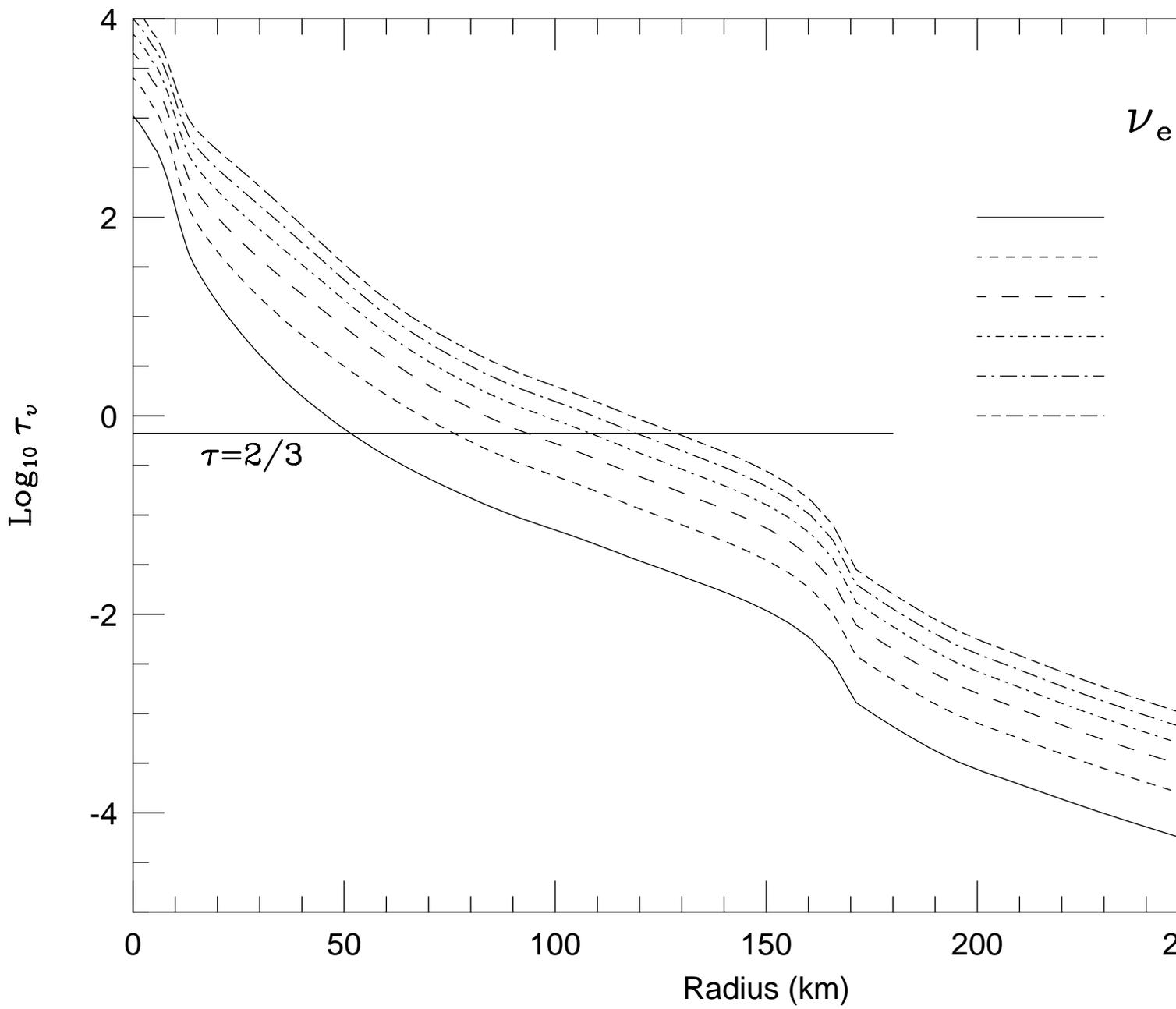
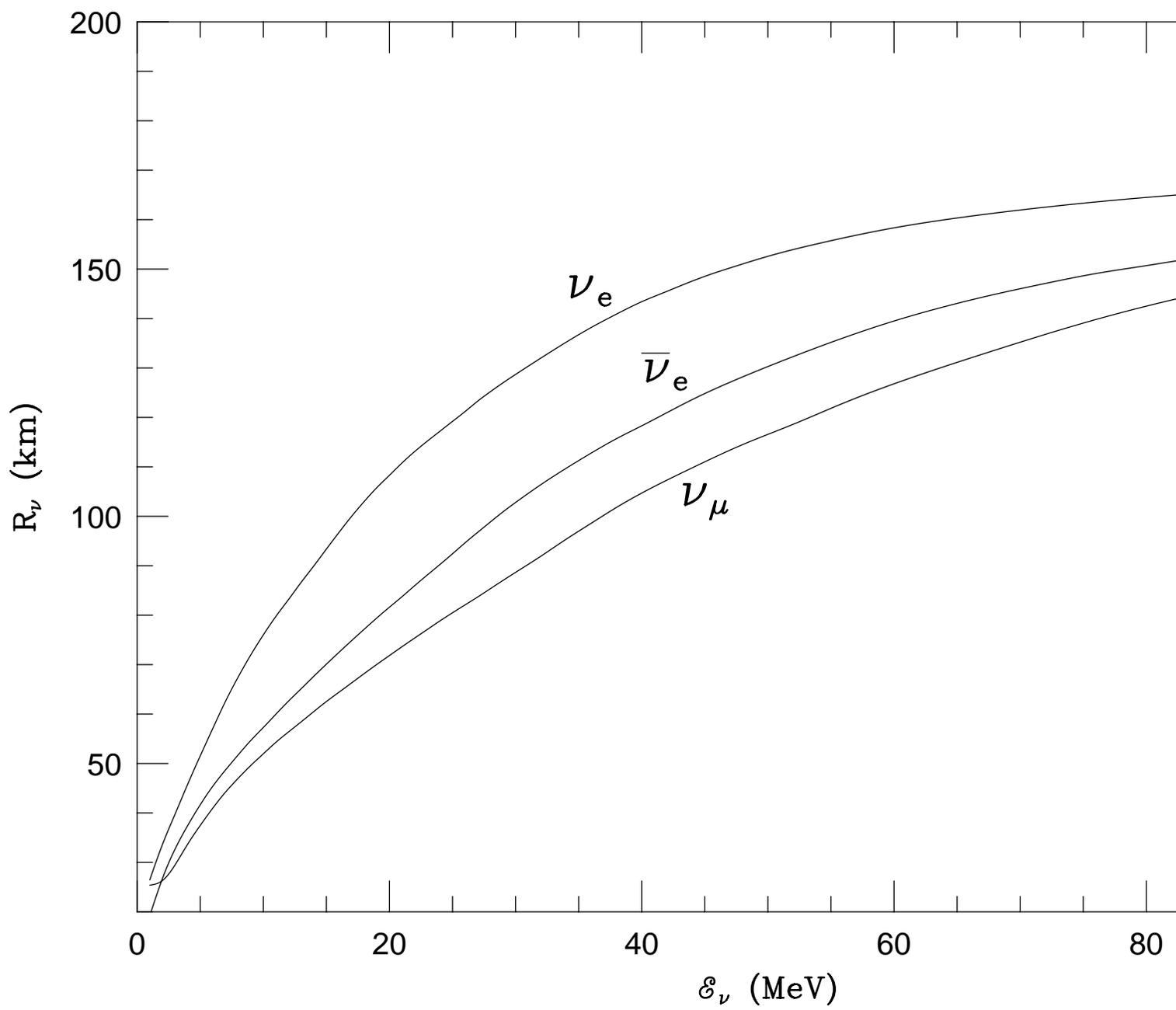
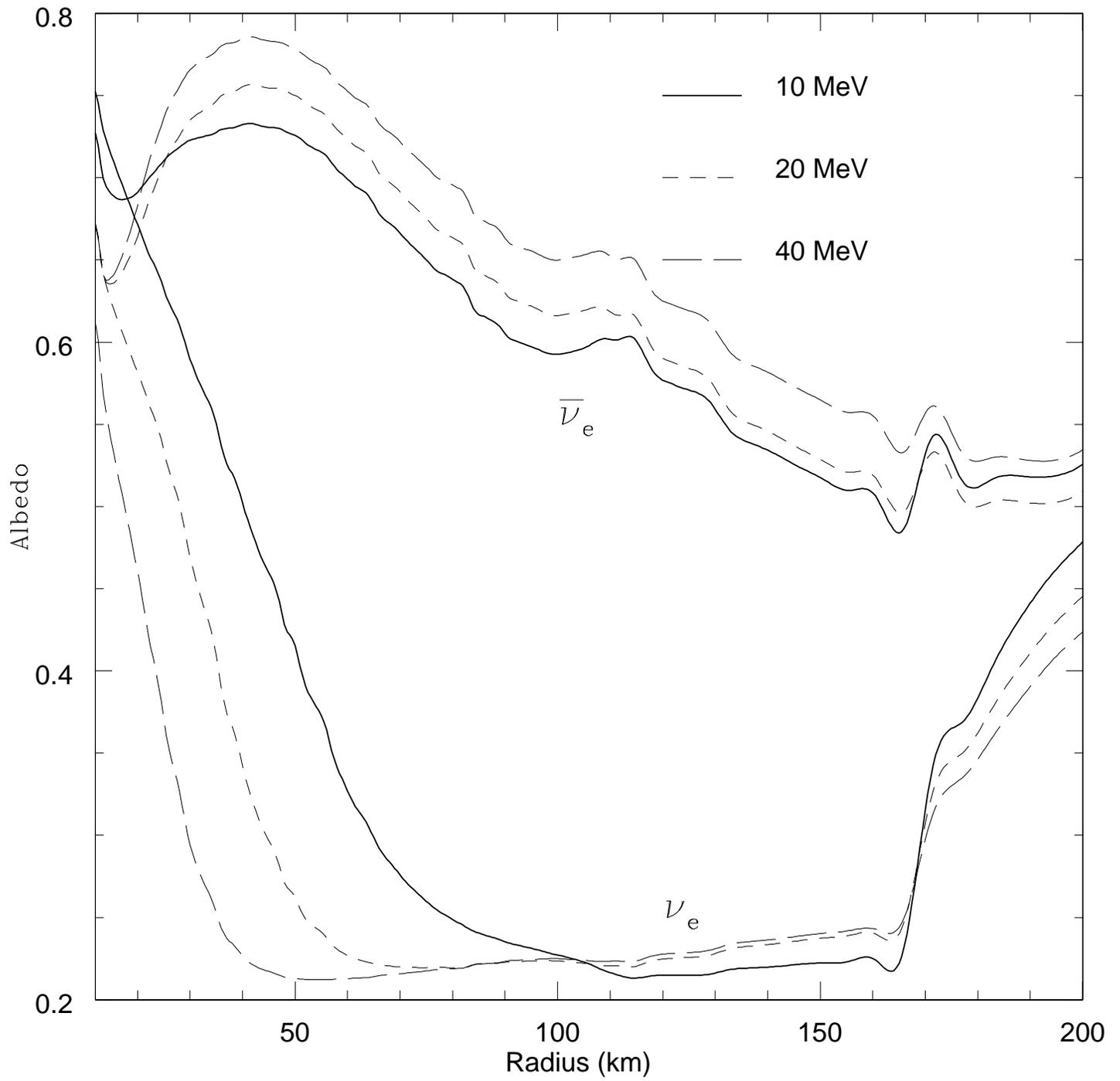
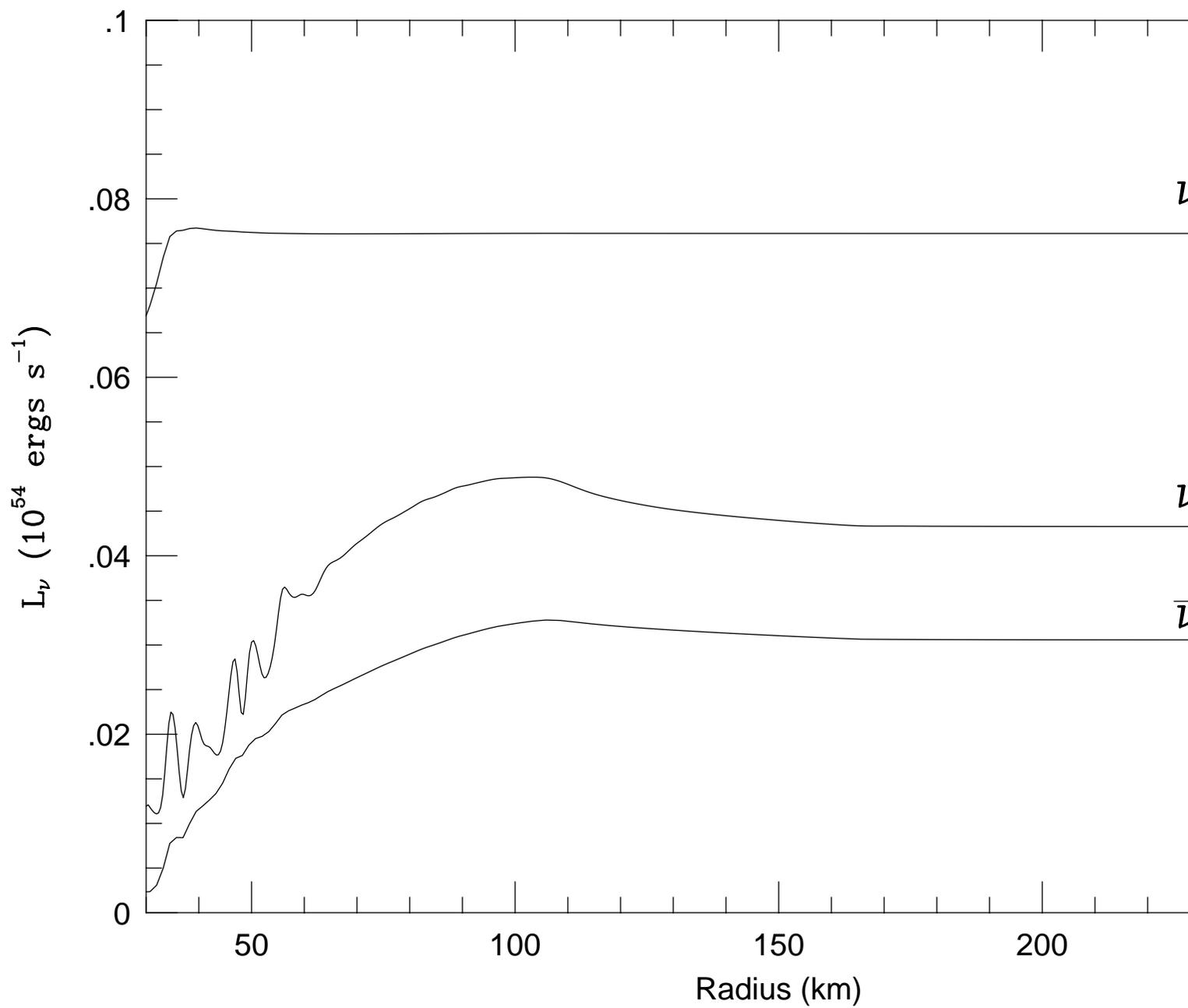
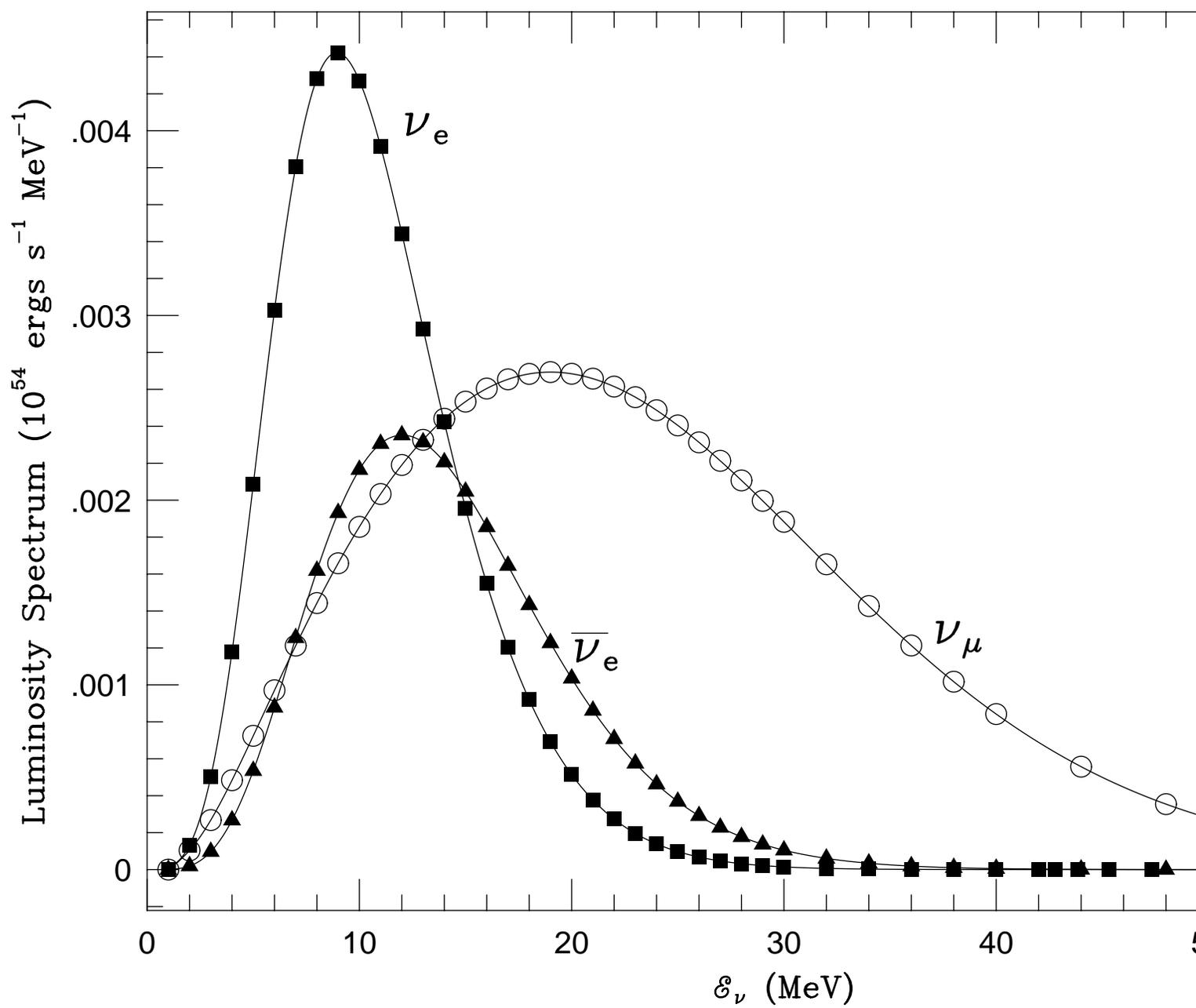
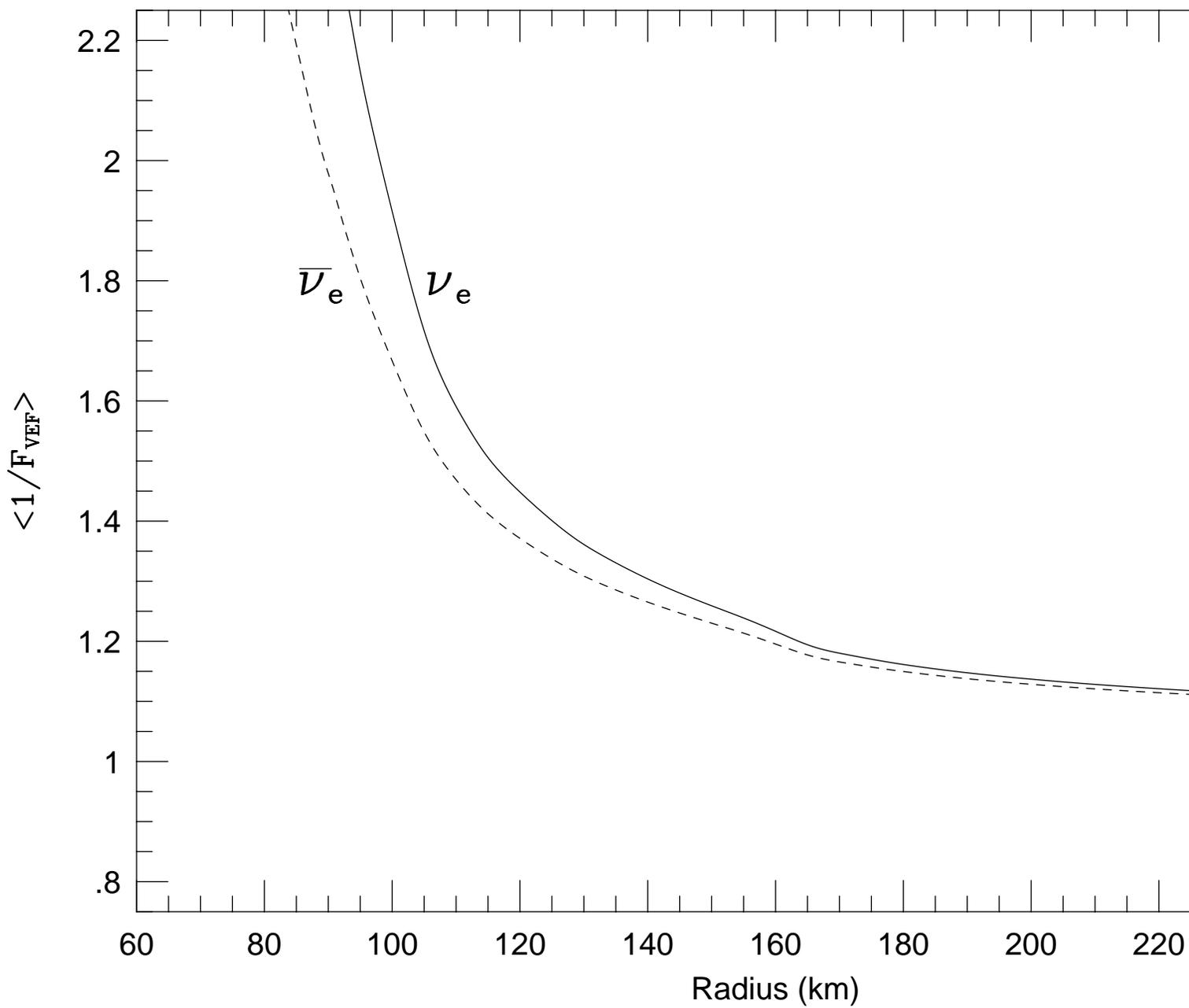
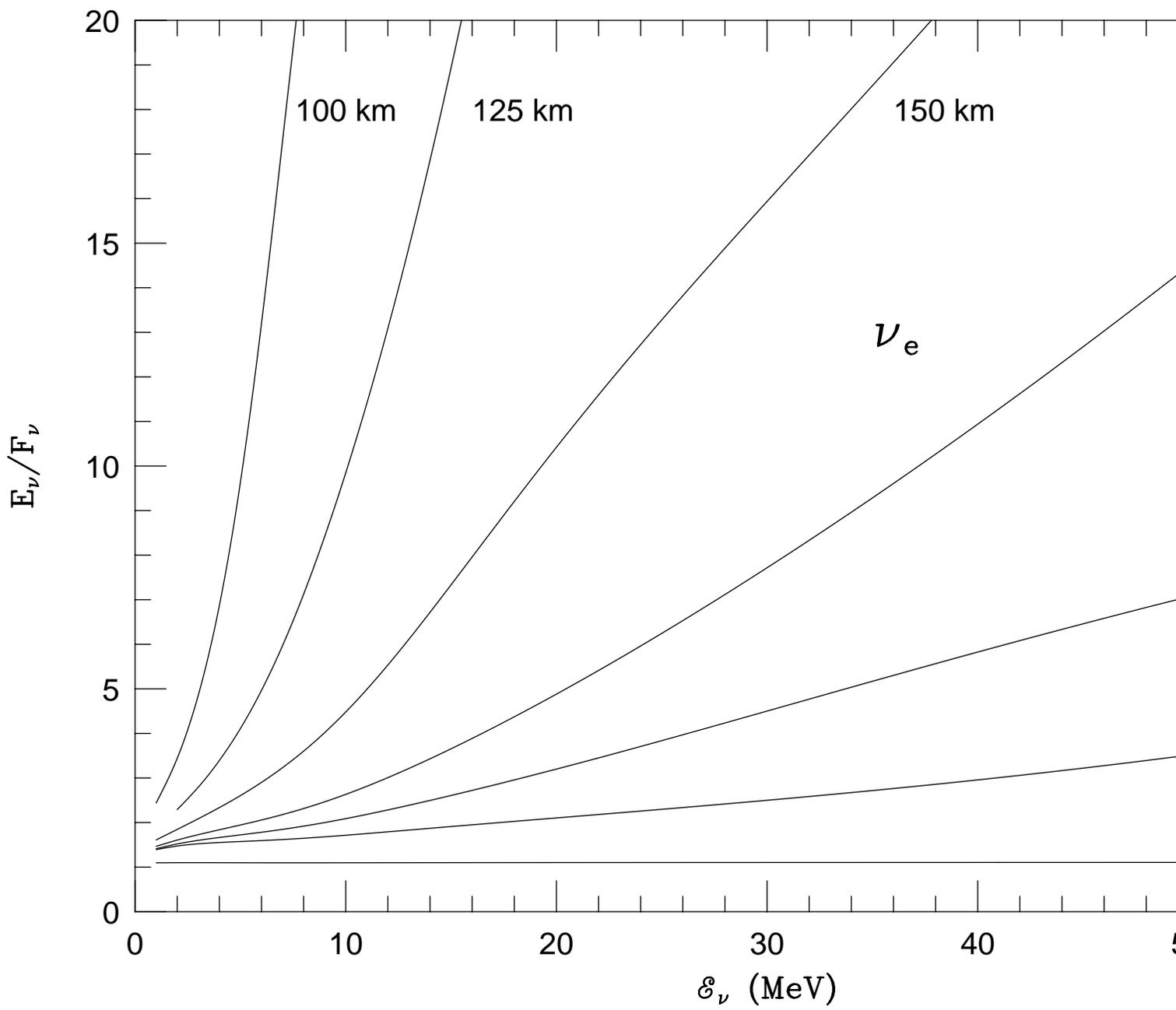
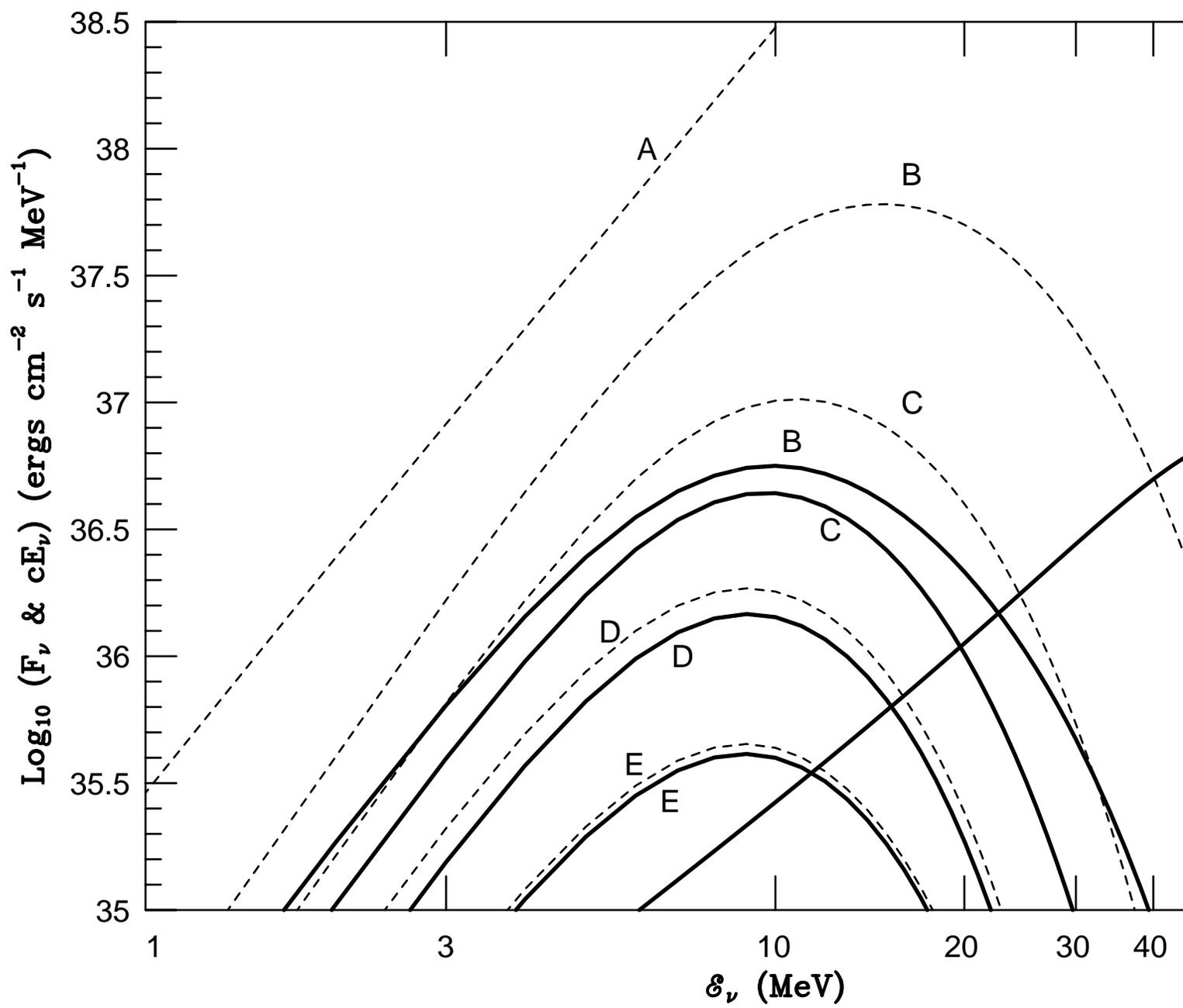
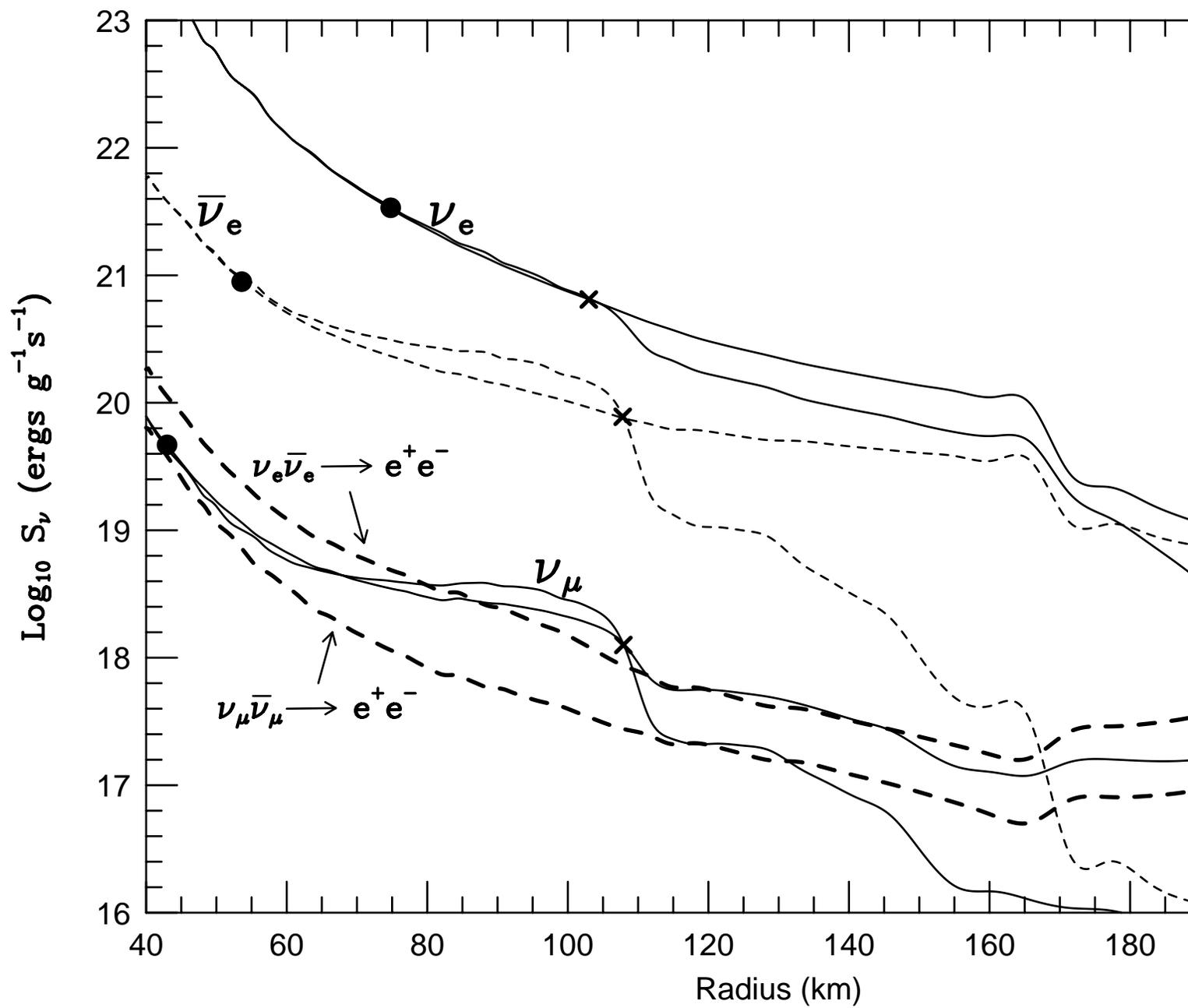
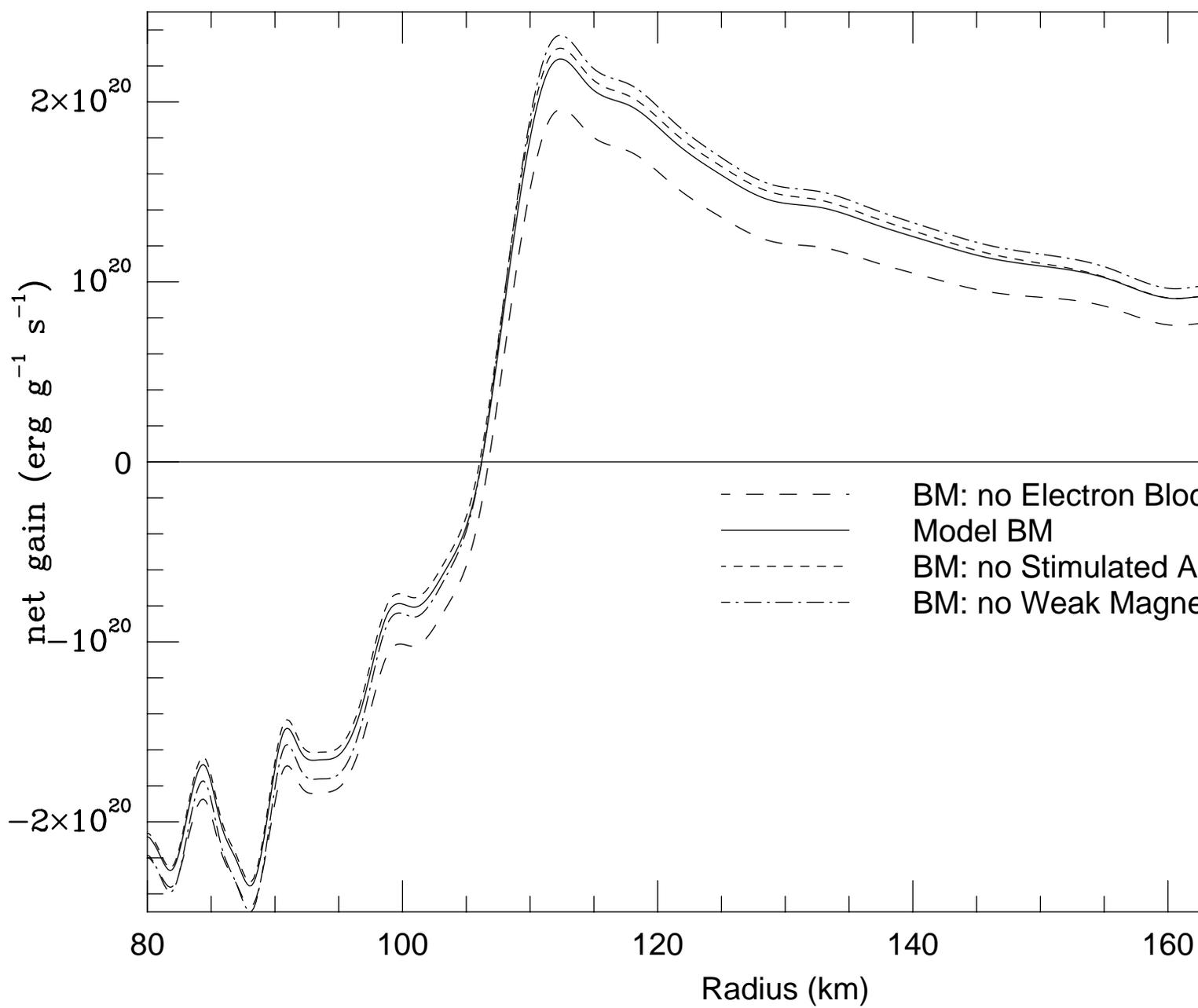
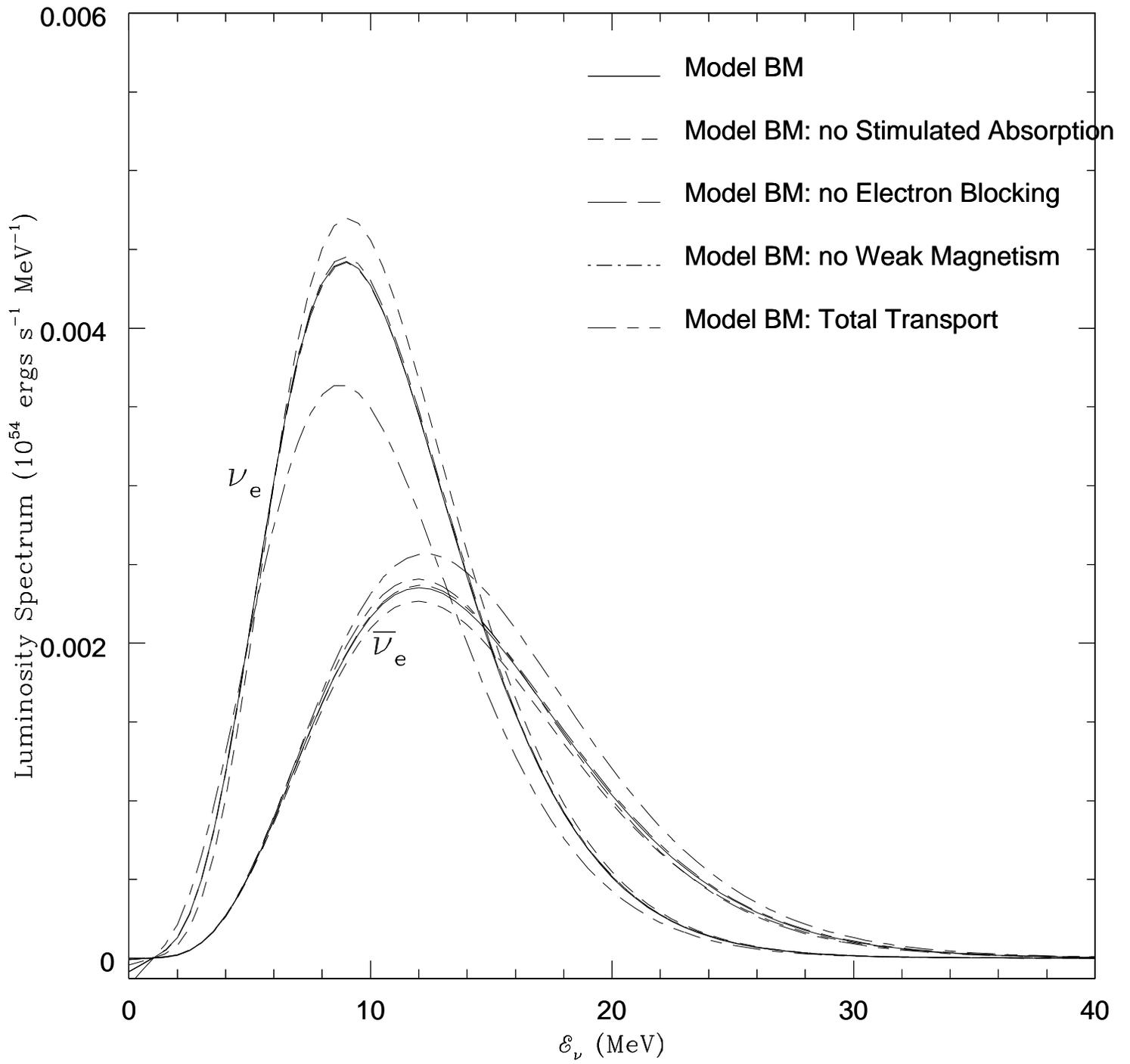
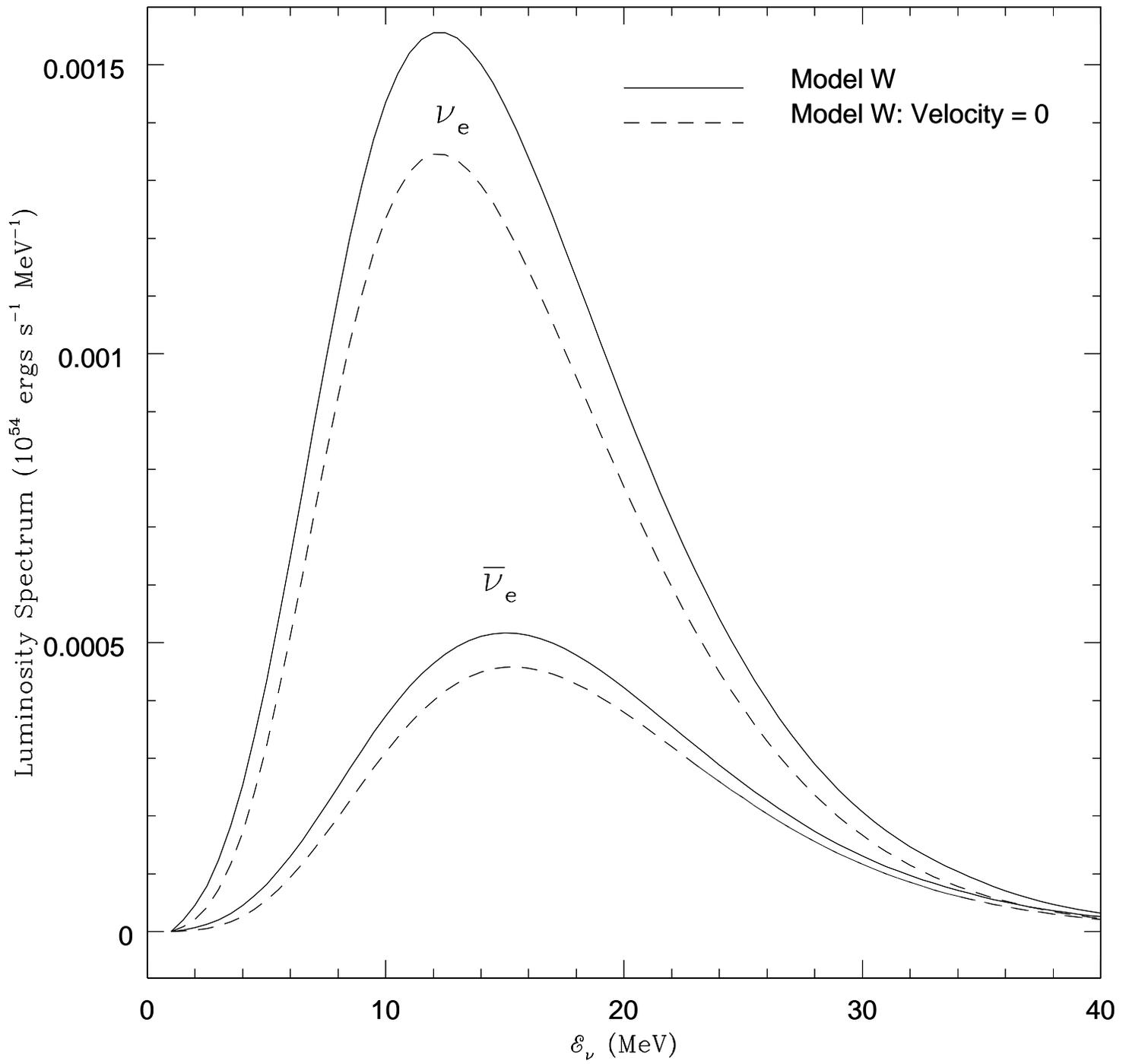
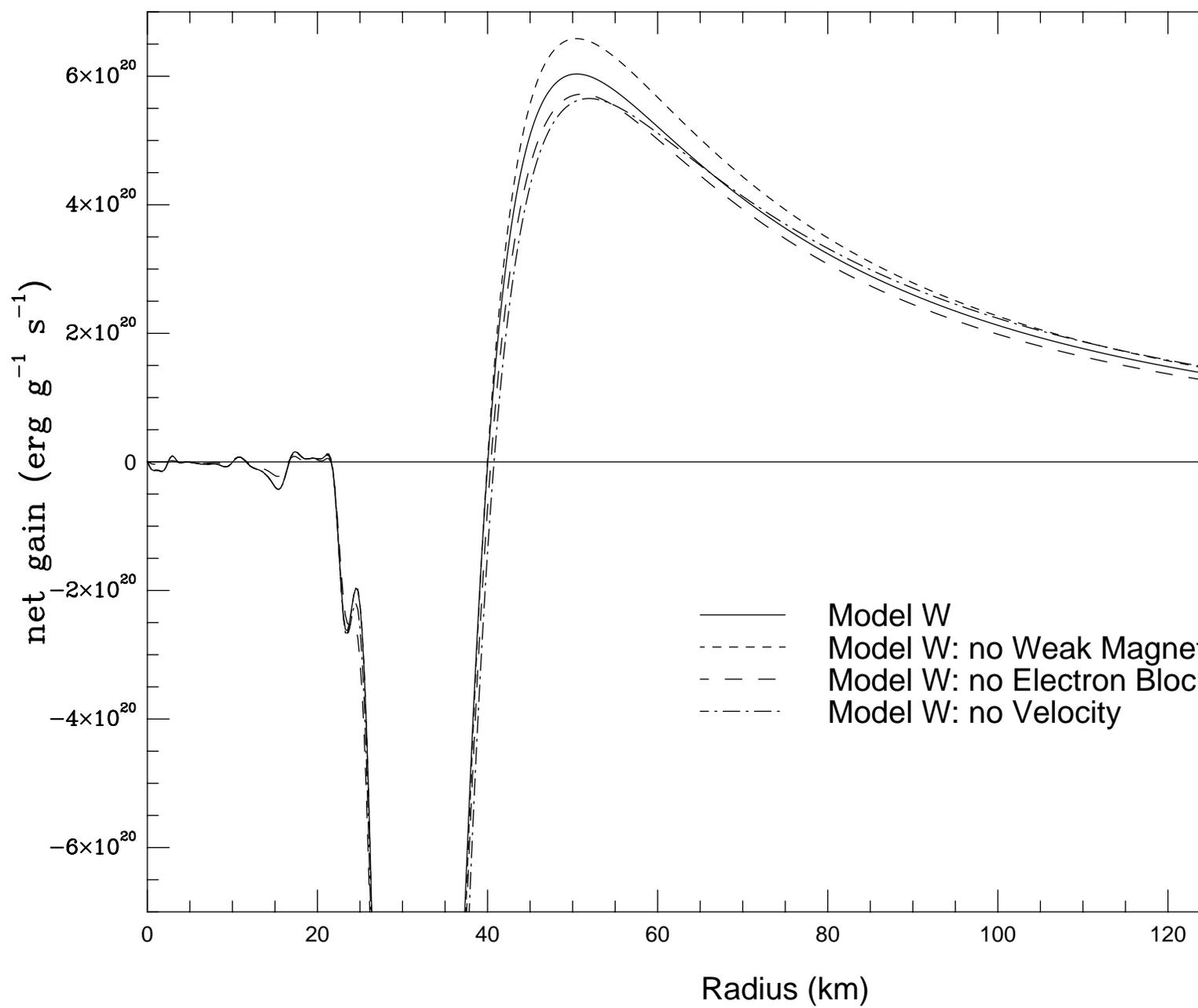
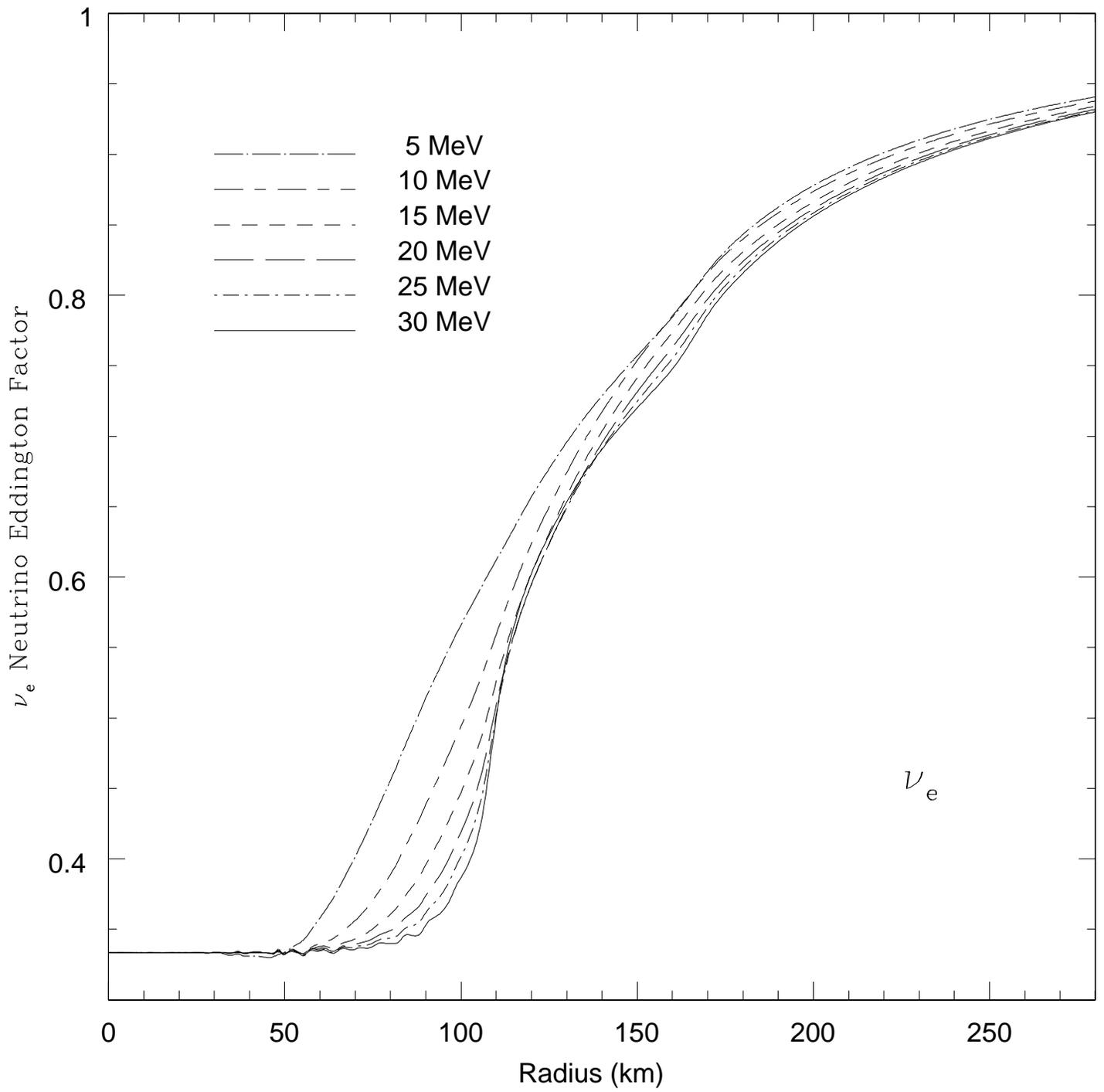
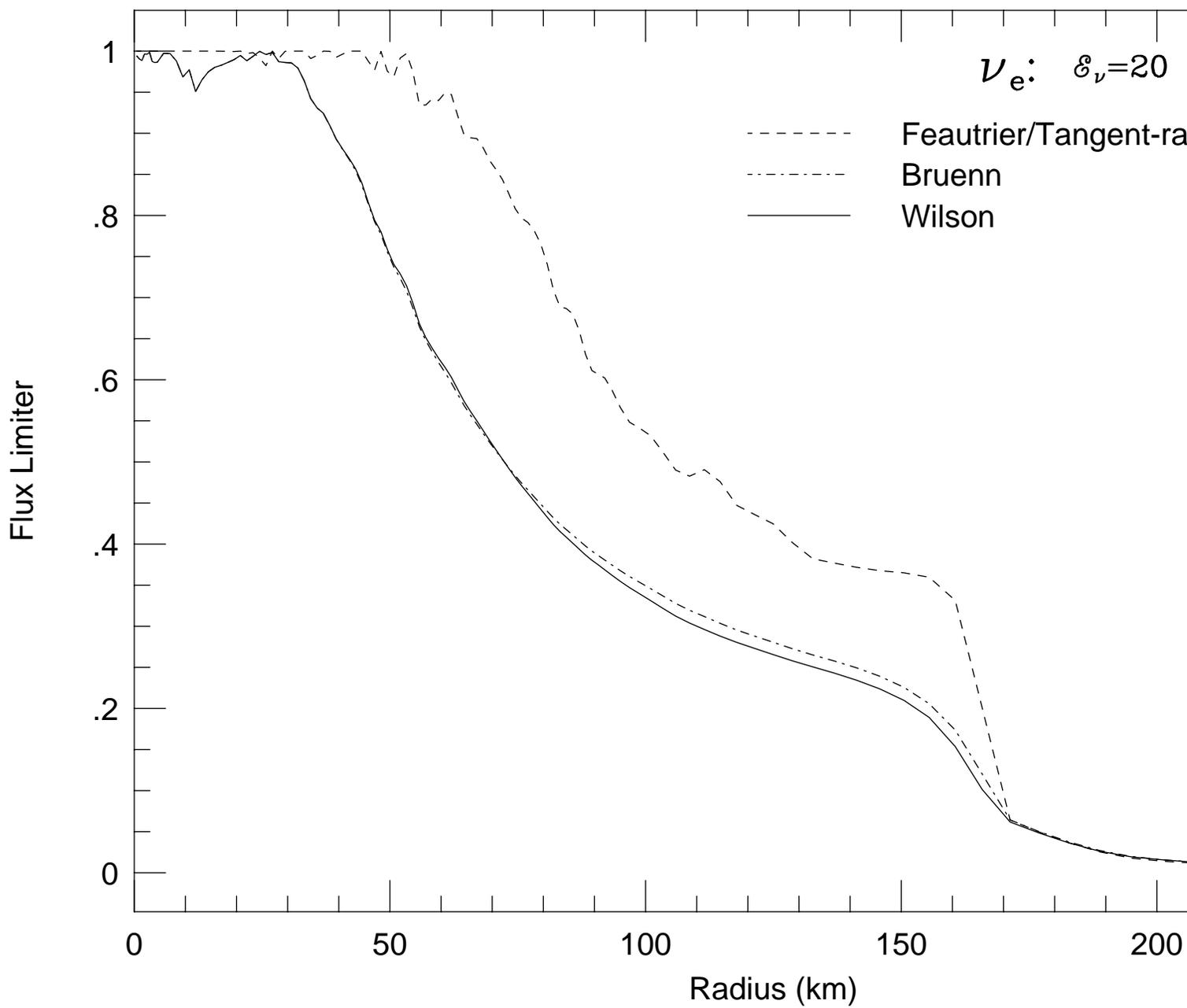
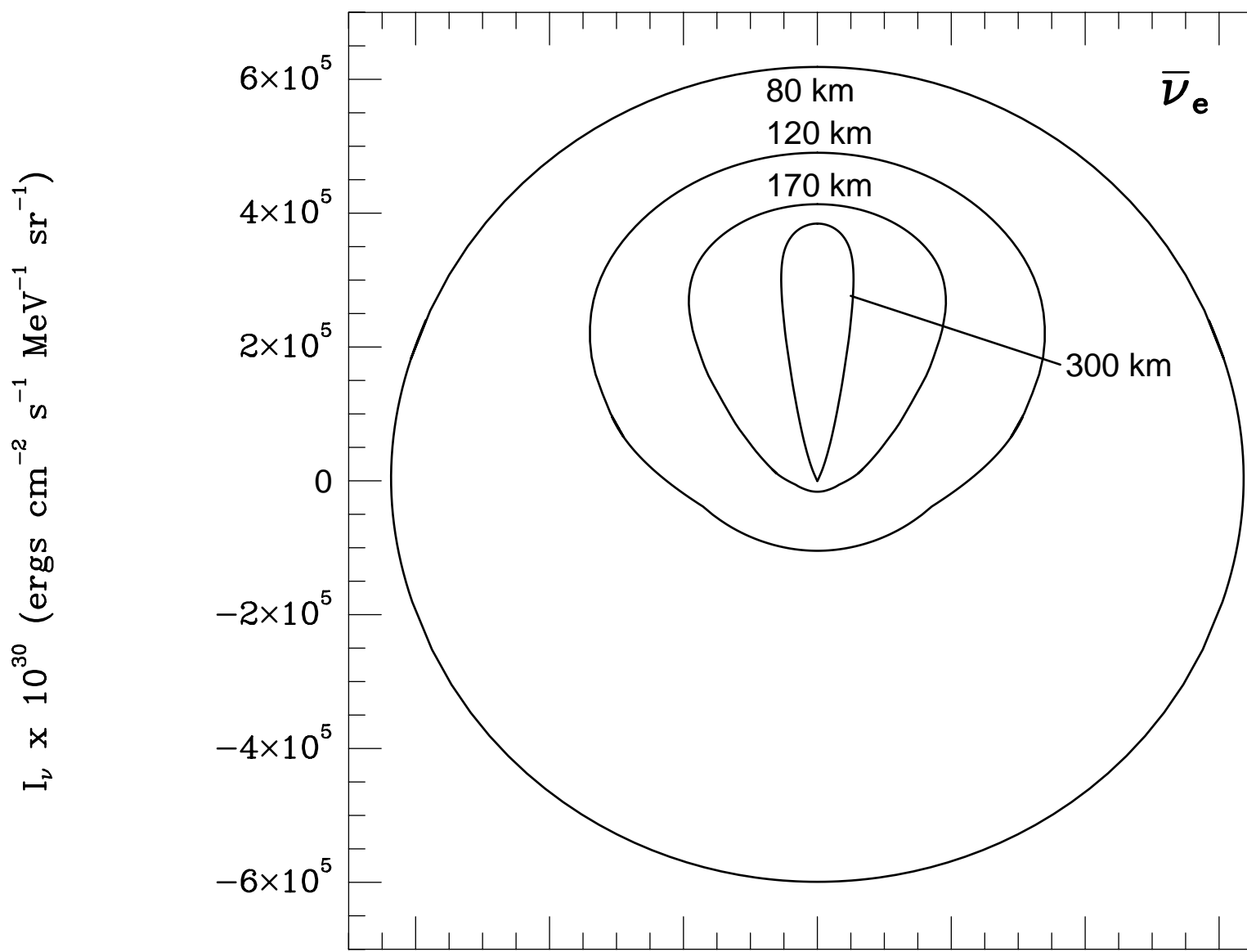
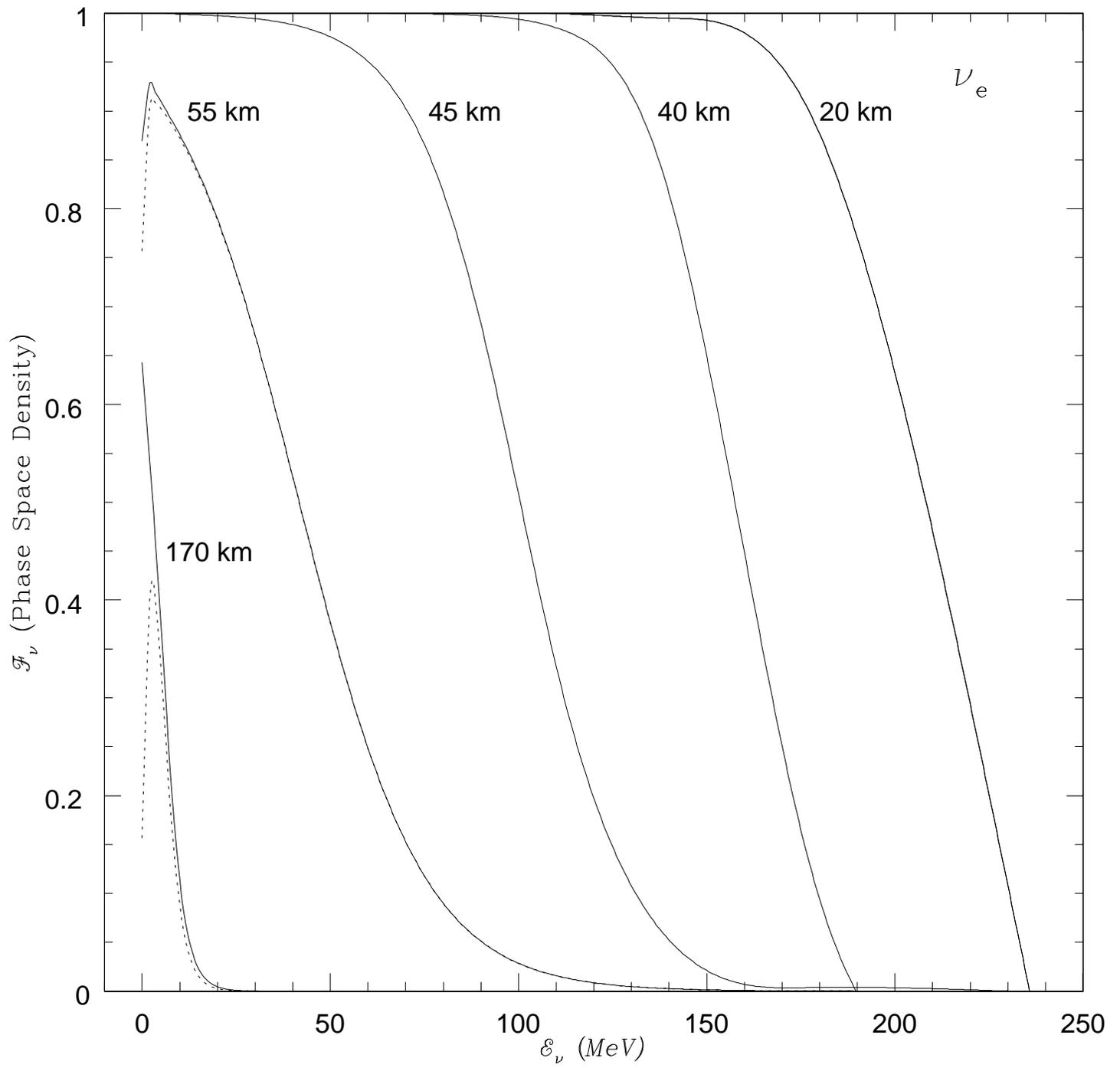
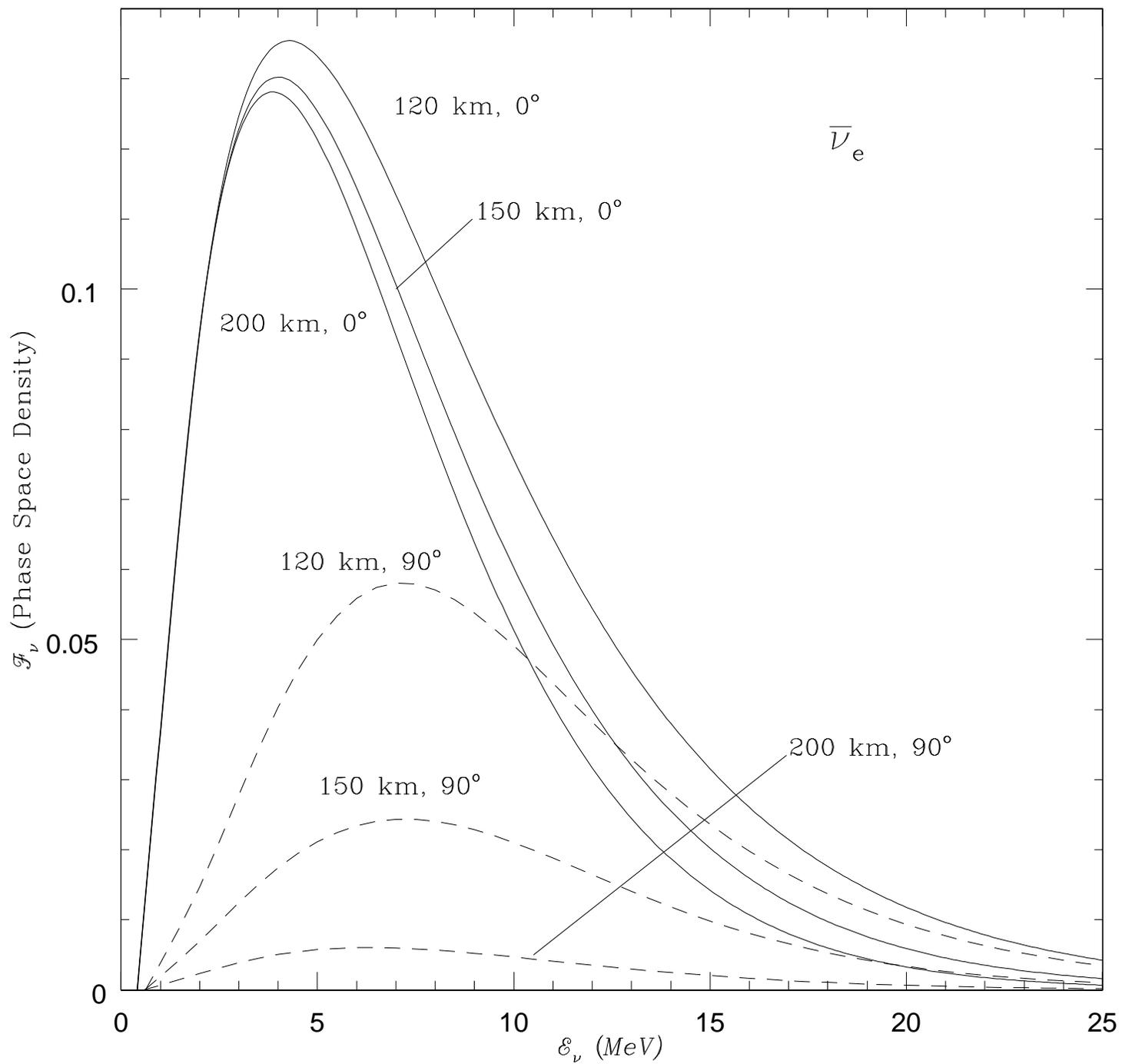
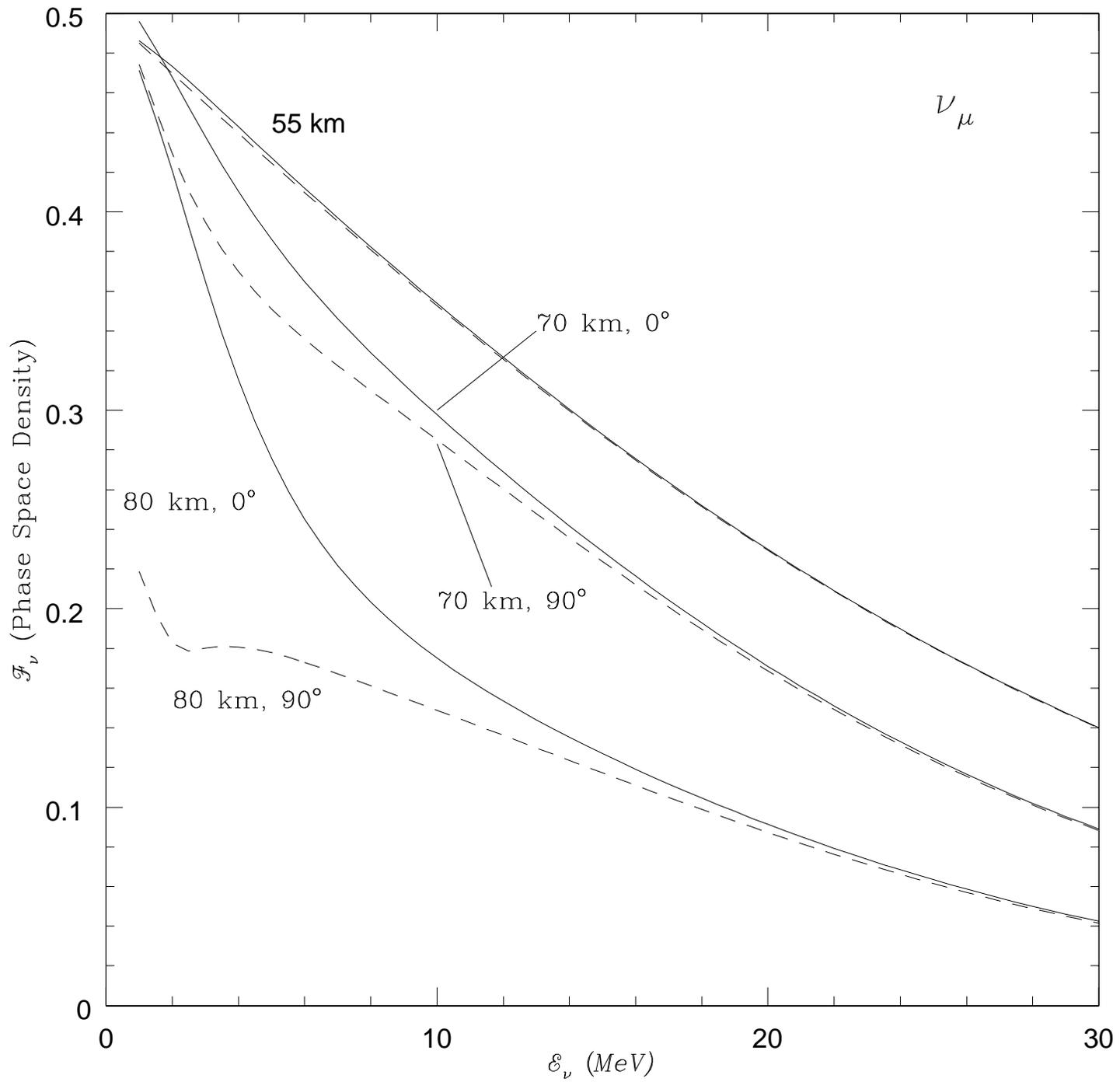
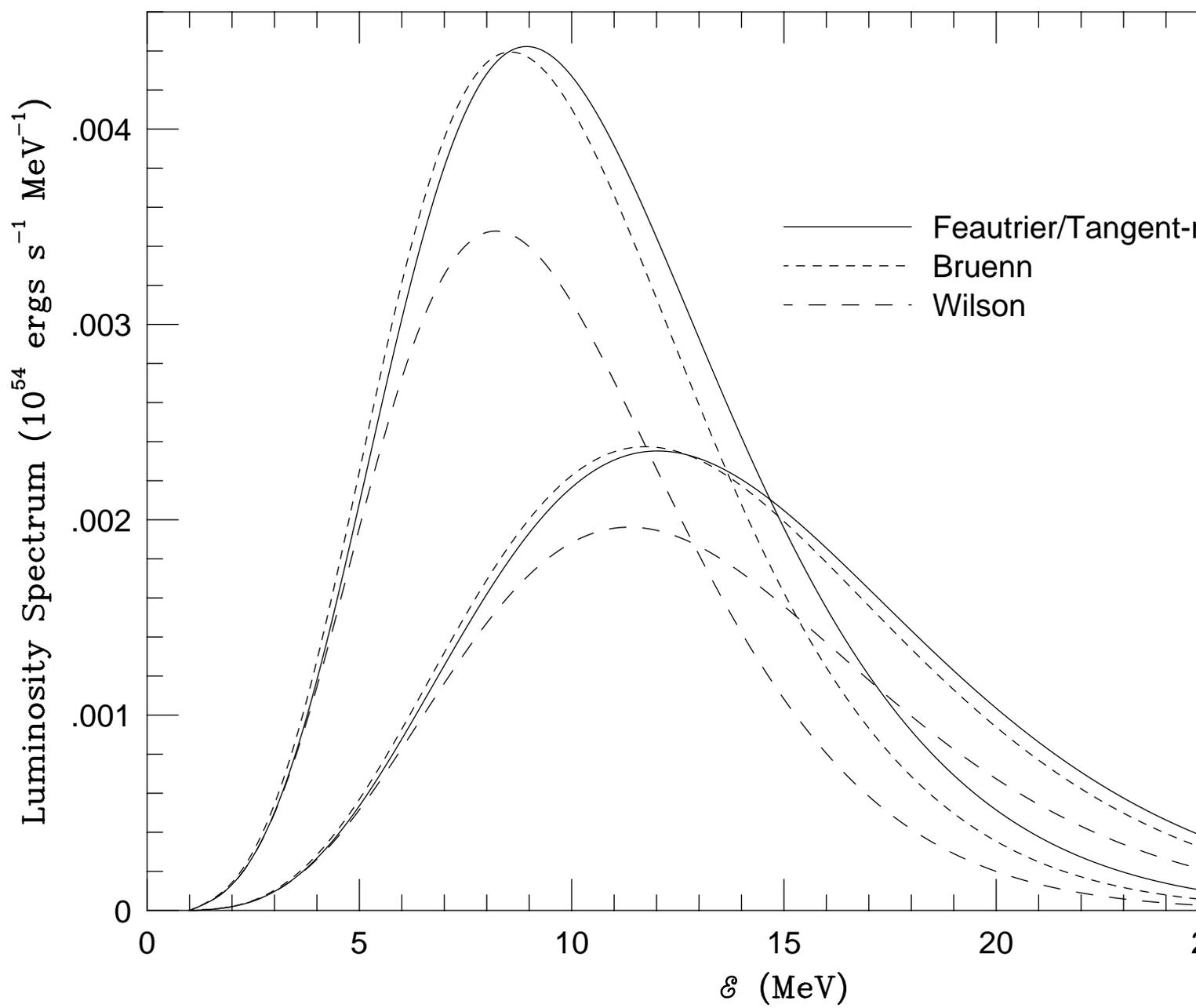
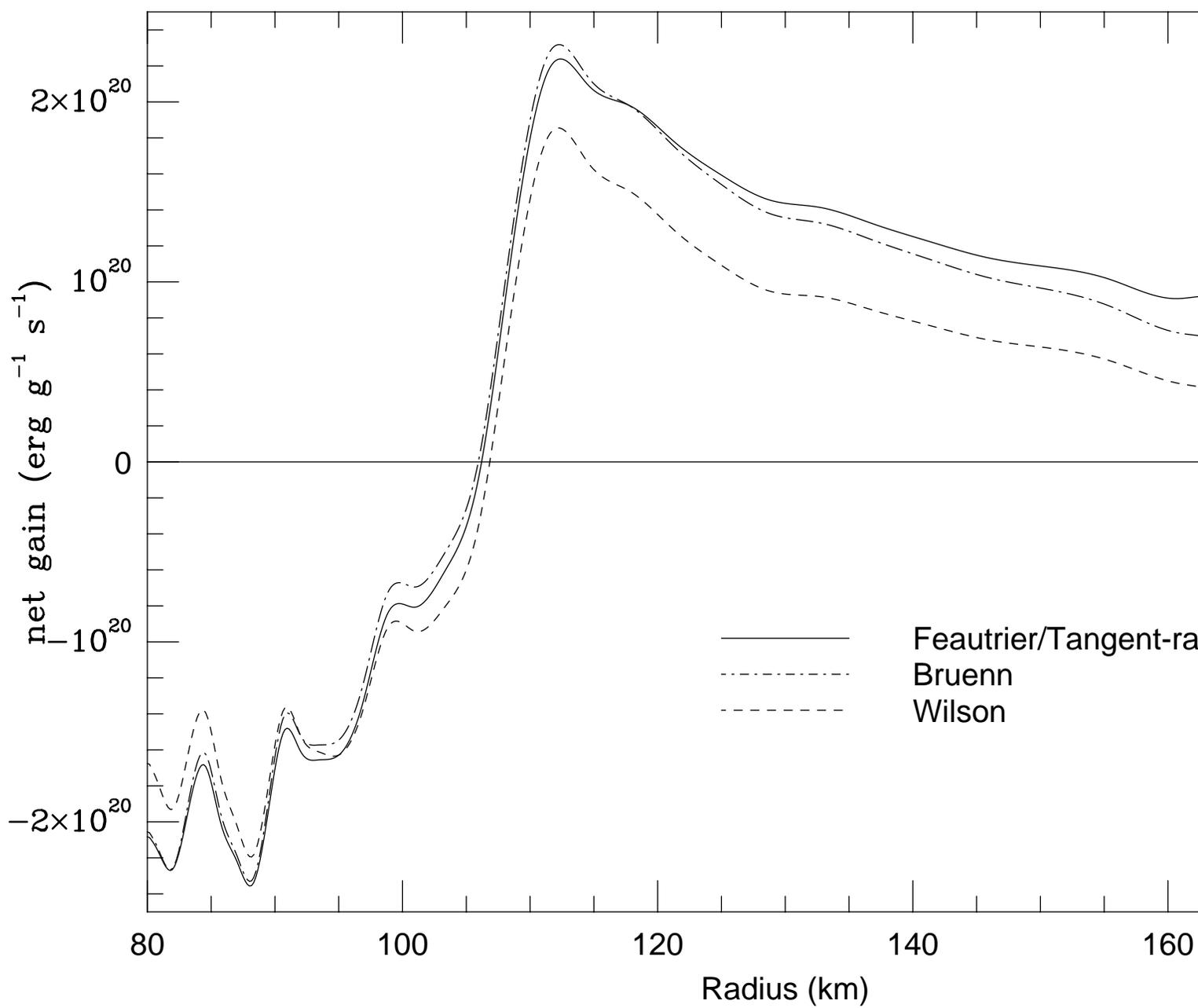
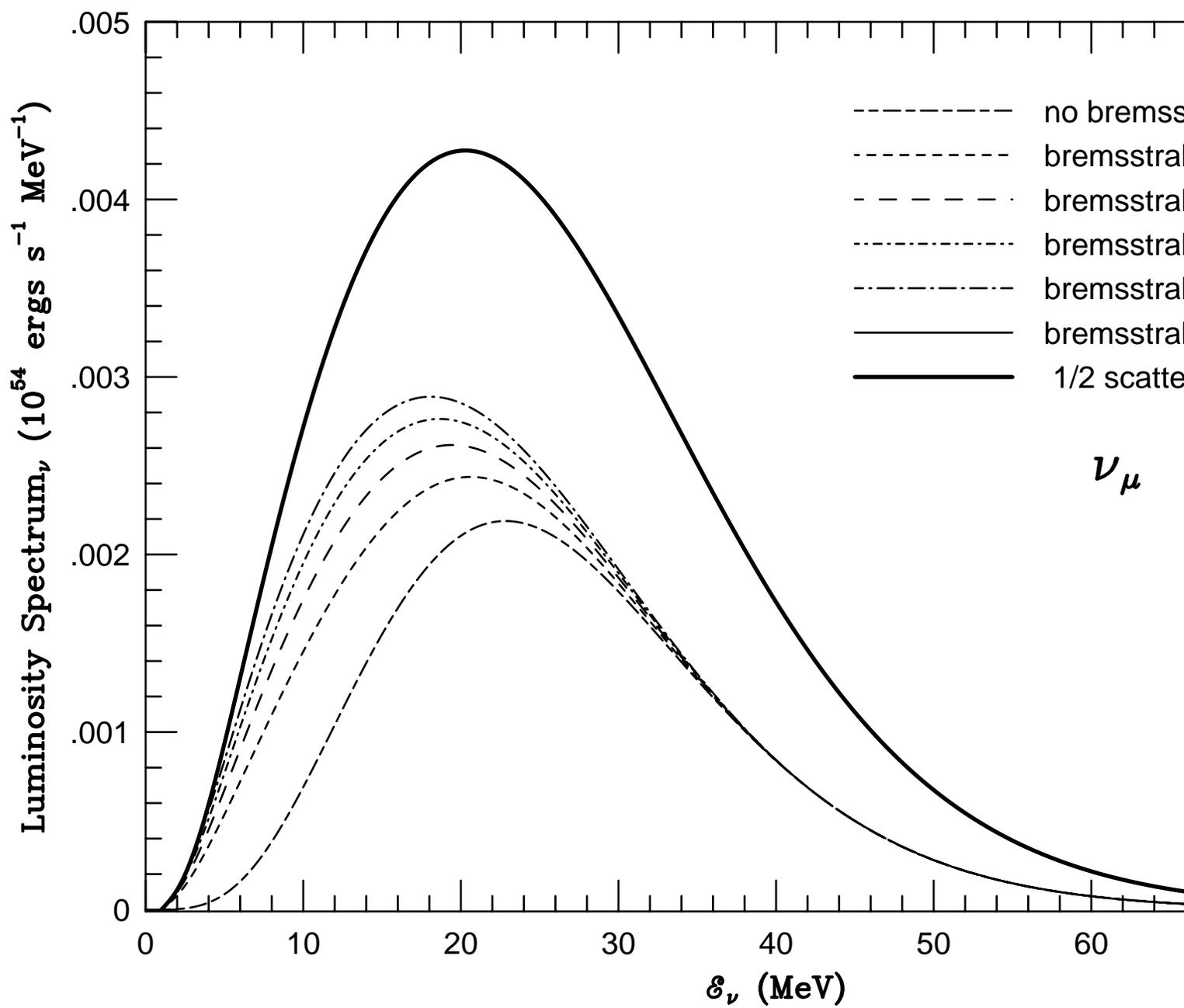
\end{document}